\documentclass[10pt,journal,cspaper,compsoc]{IEEEtran}
\ifCLASSOPTIONcompsoc
\else
\fi
\ifCLASSINFOpdf
\else
\fi

\usepackage{graphicx} 
\usepackage{subfigure}
\usepackage{algorithm}
\usepackage{algorithmic}
\usepackage{hyperref}
\usepackage{amsmath,  amssymb, amsthm}
\usepackage{cite}

\usepackage{color}

\newcommand{\beq}{\vspace{0mm}\begin{equation}}
\newcommand{\eeq}{\vspace{0mm}\end{equation}}
\newcommand{\beqs}{\vspace{0mm}\begin{eqnarray}}
\newcommand{\eeqs}{\vspace{0mm}\end{eqnarray}}
\newcommand{\barr}{\begin{array}}
\newcommand{\earr}{\end{array}}

\newcommand{\Mmat}{{\bf M}}

\newcommand{\nv}[0]{{\boldsymbol{n}}}

\newcommand{\rv}{\boldsymbol{r}}

\newcommand{\xv}{\boldsymbol{x}}

\newcommand{\cdotv}{\boldsymbol{\cdot}}

\newcommand{\Thetamat}{\boldsymbol{\Theta}}

\newcommand{\Phimat}{\boldsymbol{\Phi}}

\newcommand{\thetav}{\boldsymbol{\theta}}

\newcommand{\phiv}{\boldsymbol{\phi}}

\newcommand{\R}{\mathbb{R}}
\newcommand{\E}{\mathbb{E}}

\newtheorem{thm}{Theorem} 
\newtheorem{cor}[thm]{Corollary}

\begin{document}
%
\title{Negative Binomial Process\linebreak Count and Mixture Modeling}
%
%
%
%

\author{Mingyuan~Zhou 
                and Lawrence~Carin,~\IEEEmembership{Fellow,~IEEE}
\IEEEcompsocitemizethanks{\IEEEcompsocthanksitem M. Zhou 
is with the Department of Information, Risk, and Operations Management, McCombs School of Business, 
University of Texas at Austin, Austin, TX 78712. L. Carin is with the Department
of Electrical and Computer Engineering, Duke University, Durham, NC 27708.}
\thanks{}}

\markboth{
}%
{Shell \MakeLowercase{\textit{et al.}}: Bare Demo of IEEEtran.cls for Computer Society Journals}
%


\IEEEcompsoctitleabstractindextext{%
\begin{abstract}
The seemingly disjoint problems of count and mixture modeling are united under the  negative binomial (NB) process. 
A gamma process is employed to model the rate measure of a Poisson process, whose normalization
provides a random probability measure for mixture modeling and whose marginalization 
leads to an NB process for count modeling. A draw from the NB process 
consists of a Poisson distributed
finite number of distinct atoms, each of which is associated with a logarithmic distributed number of data samples. We reveal relationships between various count- and mixture-modeling distributions 
and construct a Poisson-logarithmic bivariate distribution that connects the NB and Chinese restaurant table distributions.
Fundamental properties of the models are developed, and we derive efficient Bayesian inference.
It is shown that with augmentation and normalization, the NB process and gamma-NB process can be reduced to the Dirichlet process and hierarchical Dirichlet process, respectively. These relationships highlight  theoretical, structural and computational advantages of the NB process. A variety of NB processes, including the beta-geometric, beta-NB, marked-beta-NB, marked-gamma-NB and zero-inflated-NB processes, with distinct sharing mechanisms, are also constructed. These models are applied to topic modeling, with connections made to existing algorithms  under 
Poisson factor analysis. 
Example results show the importance of inferring both the NB dispersion and probability parameters. 
\end{abstract}

}

\maketitle

\IEEEdisplaynotcompsoctitleabstractindextext

%
\IEEEpeerreviewmaketitle

\section{Introduction}
%
%

%
%
%
%
\IEEEPARstart{C}{ount}  data 
appear in many settings, such as 
predicting the number of motor insurance claims \cite{InvGaussianPoisson,LGNB}, analyzing infectious diseases \cite{JLS07} and modeling topics of document corpora \cite{Hofmann99probabilisticlatent,LDA,CannyGaP,DCA,BNBP_PFA}. There has been increasing interest in count modeling using the Poisson process, geometric process \cite{PoissonP,Wolpert98poisson/gammarandom,InfGaP, Thibaux,Miller} and recently the negative binomial (NB) process  \cite{BNBP_PFA,NBPJordan,MingyuanNIPS2012}. 
It is  shown in \cite{BNBP_PFA} and further demonstrated in \cite{MingyuanNIPS2012} that the NB process, originally constructed for count analysis, can be naturally applied for mixture modeling of \emph{grouped} data $\xv_1,\cdots,\xv_J$, where each group $\xv_j=\{x_{ji}\}_{i=1,N_j}$. For example, in topic modeling (mixed-membership modeling), each document consists of  a group of exchangeable words and each word is a group member that is assigned to a topic; the number of times a topic appears in a document is a latent count random variable that could be well modeled with an NB distribution \cite{BNBP_PFA,MingyuanNIPS2012}.

Mixture modeling, which infers random probability measures to assign data samples into clusters (mixture components), is a key research area of statistics and machine learning.
Although the number of samples assigned to clusters are counts, 
mixture modeling is not typically considered as a count-modeling problem. 
 It is often addressed under the Dirichlet-multinomial framework,   using the Dirichlet process   \cite{ferguson73,DP_Mixture_Antoniak,Escobar1995,MullerDP,NealDPM,Teh2010a} as the prior distribution.
 With the Dirichlet-multinomial
conjugacy, 
the Dirichlet process mixture model enjoys tractability because the posterior of the random probability measure is still a Dirichlet process.
Despite its popularity,  the Dirichlet process is inflexible in that a single concentration parameter controls both the variability of the mass around the mean \cite{Teh2010a,Wolp:Clyd:Tu:2011} and the distribution of the
number of distinct atoms  \cite{Escobar1995,lijoi2007controlling}.   For mixture modeling of grouped data, the hierarchical Dirichlet process (HDP) \cite{HDP} has been further proposed to share statistical strength between groups. The HDP inherits the same inflexibility of the Dirichlet process, and due to the non-conjugacy between Dirichlet processes, its inference  has to be solved under alternative constructions, such as the Chinese restaurant franchise and stick-breaking representations \cite{HDP,HDP-HMM,VBHDP}. To make the number of distinct atoms increase at 
  a rate faster than that of 
  the Dirichlet process, 
  one may consider the Pitman-Yor process \cite{pitman1997two, ishwaran2001gibbs} or the normalized generalized gamma process \cite{lijoi2007controlling} that provide extra parameters to add flexibility. 

 To construct more expressive mixture models with 
tractable inference, in this paper we consider mixture modeling as a count-modeling problem. Directly modeling the counts assigned to mixture components as NB random variables, we perform joint count and mixture modeling via the NB process, using completely random measures \cite{Kingman,PoissonP,JordanCRMbook,Wolp:Clyd:Tu:2011} that are easy to construct and amenable to posterior computation.
By constructing a bivariate count distribution that connects the Poisson, logarithmic, NB and Chinese restaurant table distributions,
we develop data augmentation and marginalization techniques unique to the NB distribution, with which we augment an NB process into both the gamma-Poisson and compound Poisson representations, yielding unification of  count and mixture modeling, derivation of fundamental model properties, as well as  efficient Bayesian inference.

Under the NB process, 
we employ a gamma process to  model the rate measure of a Poisson process. The normalization of the gamma process provides a random probability measure (not necessarily a Dirichlet process) for mixture modeling, and the marginalization of the gamma process leads to an NB process for count modeling. Since the gamma scale parameters appear as NB probability parameters when the gamma processes are marginalized out,
they 
  directly control count distributions on atoms and  
  they could be conveniently inferred with the beta-NB conjugacy.
    For mixture modeling of grouped data, we construct hierarchical models 
    by 
    employing an NB process for each group and sharing their NB dispersion or probability measures across groups. 
        Different parameterizations of the NB dispersion and probability parameters 
     result in a wide variety of NB processes, which are connected to previously proposed nonparametric Bayesian  mixture models. 
    The proposed joint count and mixture modeling framework provides new opportunities for better data fitting,  efficient inference and flexible model constructions.

\subsection{Related Work}
Parts of the work presented here first appeared in  \cite{BNBP_PFA,LGNB,MingyuanNIPS2012}. In this paper, we unify related materials scattered in these three conference papers and provide significant expansions. In particular, we construct a Poisson-logarithmic bivariate distribution that tightly connects the NB and Chinese restaurant table distributions, extending the Chinese restaurant process to describe the case that both the numbers of customers and tables are random variables, and we provide necessary conditions to recover the NB process and the gamma-NB process from the Dirichlet process and HDP, respectively.

We mention that a related beta-NB process has been independently
investigated 
in \cite{NBPJordan}. 
Our constructions of a wide variety of NB processes, including the beta-NB processes in \cite{BNBP_PFA} and \cite{NBPJordan} as special cases, are built on our thorough investigation of the properties, relationships and inference of the NB and related stochastic processes. In particular, we show that the gamma-Poisson construction of the NB process is key to uniting count and mixture modeling, and there are two equivalent augmentations of the NB process that allow us to develop analytic conditional posteriors and predictive distributions. These insights are not provided 
in \cite{NBPJordan}, and 
the NB dispersion parameters there are  empirically set  rather than inferred. More distinctions will be discussed along with specific models.

The remainder of the paper is organized as follows.  We review 
 some  commonly used nonparametric Bayesian priors in Section \ref{sec:preliminaries} and study the NB distribution in Section \ref{sec:CountDist}. 
 We present the NB process in Section \ref{sec:PP}, the gamma-NB process in Section \ref{sec:GNBP}, and the NB process family in  Section \ref{sec:BNB}.
  We discuss NB process topic modeling in Section \ref{sec:PFA} and present example results in Section \ref{sec:Results}. 
  
%

\section{Preliminaries}\label{sec:preliminaries}

\subsection{Completely Random Measures}

Following \cite{Wolp:Clyd:Tu:2011}, for any $\nu^+\ge0$ and any probability distribution $\pi(dp d\omega)$ on the product space $\mathbb{R}\times\Omega$, let $K^+\sim \mbox{Pois}(\nu^+)$ and ${(p_k,\omega_k)} \stackrel{iid}{\sim} \pi(dp d\omega)$ for $k=1,\cdots,K^+$. Defining $\mathbf{1}_A(\omega_k)$ as being one if $\omega_k\in A$ and zero otherwise, the random measure  $\mathcal{L}(A)\equiv\sum_{k=1}^{K^+} \mathbf{1}_A(\omega_k)p_k$ assigns independent infinitely divisible random variables $\mathcal{L}(A_i)$ to disjoint Borel sets $A_i\subset\Omega$, with characteristic functions
\beqs\label{LevyCF}
&\E\big[e^{it\mathcal{L}(A)}\big]=\exp\left\{\int\int_{\mathbb{R}\times A}(e^{itp}-1)\nu(dp d\omega)\right\},
\eeqs
where $\nu(dp d\omega)\equiv \nu^+ \pi(dp d\omega)$. A random signed measure $\mathcal{L}$ satisfying (\ref{LevyCF}) is called a L\'{e}vy random measure. More generally, if the L\'{e}vy measure $\nu({dp d\omega})$  satisfies 
\beqs \label{eq:LevyL1Condition}
&\int\int_{\mathbb{R}\times S}\min\{1, |p|\}\nu(dp d\omega) < \infty
\eeqs
for each compact $S\subset \Omega$, the L\'{e}vy random measure $\mathcal{L}$ is well defined, even if the Poisson intensity $\nu^+$ is infinite. 
A nonnegative L\'{e}vy random measure $\mathcal{L}$ satisfying (\ref{eq:LevyL1Condition}) was called a completely random measure \cite{Kingman,PoissonP}, 
 and it was introduced to machine
learning in \cite{JordanBP} and \cite{JordanCRMbook}. 

\subsubsection{Poisson Process}
Define a Poisson process $X\sim\mbox{PP}(G_0)$ on the product space $\mathbb{Z}_{+}\times \Omega$, where $\mathbb{Z}_{+}=\{0,1,\cdots\}$,
with a finite and continuous base measure $G_0$ over $\Omega$, such that $X(A)\sim\mbox{Pois}(G_0(A))$ for each subset $A\subset \Omega$. The L\'{e}vy measure of the Poisson process can be derived from~(\ref{LevyCF}) as
$
\nu(d u d\omega) = \delta_1(du) G_0(d\omega)
$, where $\delta_1(du)$ is a unit point mass at $u=1$. If $G_0$ is discrete (atomic) as $G_0=\sum_{k} \lambda_k \delta_{\omega_k}$, then the Poisson process definition is still valid that $X=\sum_{k} x_k \delta_{\omega_k},~ x_k\sim\mbox{Pois}(\lambda_k)$. If $G_0$ is mixed discrete-continuous, 
then $X$ is the sum of two independent contributions. 
Except where otherwise specified, 
below we consider the base measure to be finite and continuous. 

\subsubsection{Gamma Process}
 We define a gamma process 
 \cite{Wolpert98poisson/gammarandom, Wolp:Clyd:Tu:2011} $G\sim\mbox{GaP}(c,G_0)$ on the product space $\R_{+}\times \Omega$, where $\R_{+}=\{x: x\ge0\}$, with  scale parameter $1/c$ and 
  base measure $G_0$, such that $G(A)\sim\mbox{Gamma}(G_0(A),1/c)$ 
 for each subset $A\subset \Omega$, where $\mbox{Gamma}(\lambda;a,b)=\frac{1}{\Gamma(a)b^a}\lambda^{a-1}e^{-\frac{\lambda}{b}}$. 
 The gamma process 
 is a completely random measure, 
whose L\'{e}vy measure can be derived from (\ref{LevyCF}) as
$
\nu(dr d\omega) = r^{-1} e^{-cr} dr G_0(d\omega).
$
Since the Poisson intensity $\nu^{+}=\nu(\R_{+}\times\Omega)=\infty$ and $\int\int_{\R_{+}\times \Omega} r \nu(dr d\omega)$  
is finite, there are countably infinite atoms and
a draw from the gamma process 
can be expressed as
\beqs
&G = \sum_{k=1}^{\infty} r_k \delta_{\omega_k},~(r_k, \omega_k)\stackrel{iid}{\sim}  \pi(dr d\omega), \notag 
\eeqs
where $\pi(dr d\omega)\nu^{+} \equiv \nu(dr d\omega)$.

\subsubsection{Beta Process}\label{sec:preliminaries1}

 The beta process was defined by \cite{Hjort} for survival analysis with $\Omega = \R_+.$
 Thibaux and Jordan \cite{JordanBP} modified the process  by defining 
 $B\sim\mbox{BP}(c,B_0)$ on the product space $[0,1]\times\Omega$, with L\'{e}vy measure
$
 \nu(dpd\omega) = cp^{-1}(1-p)^{c-1}dpB_0(d\omega),
 $ 
 where $c>0$ is a concentration parameter 
 and $B_0$ is a base measure. 
 Since the Poisson intensity $\nu^{+}=\nu([0,1]\times\Omega)=\infty$ and $\int\int_{[0,1]\times \Omega} p \nu(dp d\omega)$  
is finite, there are countably infinite atoms and
a draw from the beta process 
can be expressed as
\beqs
&B = \sum_{k=1}^{\infty} p_k \delta_{\omega_k},~(p_k, \omega_k)\stackrel{iid}{\sim}  \pi(dp d\omega), \notag 
\eeqs
where $\pi(dp d\omega)\nu^{+} \equiv \nu(dp d\omega)$.

\subsection{Dirichlet and Chinese Restaurant Processes}
\subsubsection{Dirichlet Process}\label{sec:DP}
 Denote  $\widetilde{G} = G/G(\Omega)$, where $G\sim\mbox{GaP}(c,G_0)$, then for any measurable disjoint partition $A_1,\cdots,A_Q$ of $\Omega$, we have
 $
 \left[\widetilde{G}(A_1),\cdots,\widetilde{G}(A_Q)\right]\sim\mbox{Dir}\left(\gamma_0 \widetilde{G}_0(A_1),\cdots,\gamma_0 \widetilde{G}_0(A_Q)\right)
 $,
  where $\gamma_0=G_0(\Omega)$ 
  and $\widetilde{G}_0 = G_0/\gamma_0$. 
  Therefore, with a space invariant scale parameter $1/c$, the normalized gamma process $\widetilde{G} = G/G(\Omega)$ is a Dirichlet process \cite{ferguson73,ishwaran02} with concentration parameter $\gamma_0$ and base probability measure $\widetilde{G}_0$, expressed as $\widetilde{G} \sim \mbox{DP}(\gamma_0,\widetilde{G}_0)$.   Unlike the gamma process, the Dirichlet process is no longer a completely random measure as the random variables $\{\widetilde{G}(A_q)\}$ for disjoint sets $\{A_q\}$ are negatively correlated. 
  
  A gamma process with a space invariant scale parameter
can also be recovered from a Dirichlet process: if a gamma random variable $\alpha\sim{\mbox{Gamma}}(\gamma_0,1/c)$ and a Dirichlet process $\widetilde{G}\sim{\mbox{DP}}(\gamma_0,\widetilde{G}_0)$ are independent with
$\gamma_0=G_0(\Omega)$ and $\widetilde{G}_0 = G_0/\gamma_0$, then 
$G=\alpha\widetilde{G}$ becomes a gamma process as
$ 
G\sim{\mbox{GaP}}(c,G_0)$. 

\subsubsection{Chinese Restaurant Process}
In a Dirichlet process $\widetilde{G}\sim\mbox{DP}(\gamma_0,\widetilde{G}_0)$, 
we assume $X_i\sim \widetilde{G}$; $\{X_i\}$ are independent given $\widetilde G$ and hence exchangeable. The predictive distribution of a new data sample $X_{m+1}$, conditioning on $X_1,\cdots,X_{m}$, with $\widetilde{G}$ marginalized out, can be expressed as
\beqs\label{eq:CRP}
&X_{m+1}|X_1,\cdots,X_m \sim \E\left[\left.\widetilde{G}\right|X_1,\cdots,X_m\right]\notag\\
&=\sum_{k=1}^K \frac{n_k}{m+\gamma_0}\delta_{\omega_k} + \frac{\gamma_0}{m+\gamma_0} \widetilde{G}_0,
\eeqs
where $\{\omega_k\}_{1,K}$ are distinct atoms in $\Omega$ observed in $X_1,\cdots,X_m$ and $n_k=\sum_{i=1}^{m} X_i(\omega_k)$ is the number of data samples associated with $\omega_k$.
  The stochastic process described in (\ref{eq:CRP}) is known as the P\'{o}lya urn scheme \cite{PolyaUrn} and also the Chinese restaurant process \cite{aldous:crp,HDP,csp}.

  The number of nonempty tables  $l$ in a Chinese restaurant process, with concentration parameter $\gamma_0$ and $m$ customers, is a random variable  generated as
   $
   l= \sum_{n=1}^m b_n,~b_n\sim \mbox{{Bernoulli}}\left(\frac{\gamma_0}{n-1+\gamma_0}\right).
  $
  This random variable is referred as the Chinese restaurant table (CRT) random variable $l\sim\mbox{CRT}(m,\gamma_0)$. 
  As shown in \cite{DP_Mixture_Antoniak,Escobar1995,HDP,MingyuanNIPS2012}, it has probability mass function (PMF)
   \beqs
\hspace{-3mm}&
f_L(l|m,\gamma_0)= {\frac{\Gamma(\gamma_0)}{\Gamma(m+\gamma_0)}}|s(m,l)| \gamma_0^l, ~ l=0,1,\cdots,m\notag
\eeqs
where $s(m,l)$ are Stirling numbers of the first kind.

\section{Negative Binomial Distribution
}\label{sec:CountDist}

The Poisson distribution $m\sim \mathrm{Pois}(\lambda)$ is commonly used to model count data, with PMF
\beqs
&f_M(m) = \frac{ \lambda^m e^{-\lambda}}{m!},~m \in\mathbb{Z}_{+}.\notag
\eeqs
Its
mean and variance are both equal to $\lambda$.
Due to heterogeneity (difference between individuals) and contagion (dependence between the occurrence of events), count data are usually  overdispersed in that the variance is greater than the mean, making the Poisson assumption restrictive.
By placing a gamma 
prior with shape $r$ and scale $\frac{p}{1-p}$ on $\lambda$ as
$m\sim \mbox{Pois}(\lambda)$, $\lambda\sim\mbox{Gamma}\big(r,\frac{p}{1-p}\big)$
and marginalizing out $\lambda$, an NB distribution $m\sim\mbox{NB}(r,p)$ is obtained, with PMF
\beqs &f_M(m|r,p) =\frac{\Gamma(r+m)}{m!\Gamma(r)}(1-p)^r p^m,~m \in\mathbb{Z}_{+}, \notag
\eeqs
where $r$ is the nonnegative dispersion parameter and $p$ is the probability parameter.  Thus  the NB distribution is also known as the gamma-Poisson mixture  distribution \cite{Yule}.  It  has a mean $\mu = {rp}/(1-p)$ smaller than the variance $\sigma^2 = {rp}/{(1-p)^2} = \mu + r^{-1}\mu^2$, with the variance-to-mean ratio (VMR) as $(1-p)^{-1}$ and the overdispersion level (ODL, the coefficient of the quadratic term in $\sigma^2$) as $r^{-1}$, and thus it is usually favored over the Poisson distribution
for modeling overdispersed counts.

As shown in \cite{LogPoisNB}, $m\sim \mbox{NB}(r,p)$ can also be generated from a compound Poisson distribution as
\beqs\label{NB_CompoundPoisson}
\hspace{-3mm}&m =  \sum_{t=1}^l u_t, ~u_{t} \stackrel{iid}{\sim}  \mbox{Log}(p),~l \sim \mbox{Pois}(-r\ln(1-p)),\notag
\eeqs
where $u\sim \mbox{Log}(p)$ corresponds to the logarithmic distribution \cite{LogPoisNB,johnson2005univariate} with PMF
$
f_U(k)={-p^k}/[k\ln(1-p)],~~k=1,2,\cdots
$,
and probability-generating function (PGF)
\beqs
&C_U(z) = {\ln(1-pz)}/{\ln(1-p)},~~|z|<{p^{-1}}. \notag
\eeqs
One may also show that $\lim_{r\rightarrow \infty}\mbox{NB}(r,\frac{\lambda}{\lambda+r})=\mbox{Pois}(\lambda)$, and conditioning on $m>0$, $m\sim\mbox{NB}(r,p)$ becomes $m\sim\mbox{Log}(p)$ as ${r\rightarrow 0}$.

The NB distribution has been widely investigated and applied to numerous scientific studies  \cite{NB_Fitting_53,Cameron1998,WinkelmannCount,NB_Bio_2008}.
Although inference of the NB probability parameter $p$ is straightforward with the beta-NB conjugacy, 
inference of the NB dispersion parameter $r$, whose conjugate prior is unknown, has long been a challenge.
 The maximum likelihood (ML) approach is commonly used to estimate~$r$, however, it only provides a point estimate and does not allow 
 incorporating prior information; moreover, the ML estimator of $r$ often lacks robustness and may be severely biased or even fail to converge, especially if the sample size is small 
\cite{NB_Pieters_1977,NB_1984,LawlessNB87,ML_NB90,BiasMLE_NB,JLS07}.
 Bayesian approaches are able to model the uncertainty of estimation and incorporate prior information, however, the only available closed-form Bayesian inference for $r$ relies on approximating the ratio of two gamma functions 
 \cite{NBbayesian}.

 \subsection{Poisson-Logarithmic Bivariate Distribution}

\begin{figure}[tb]
\vskip 0.077in
\begin{center}
\centerline{\includegraphics[width=1.0\columnwidth
]{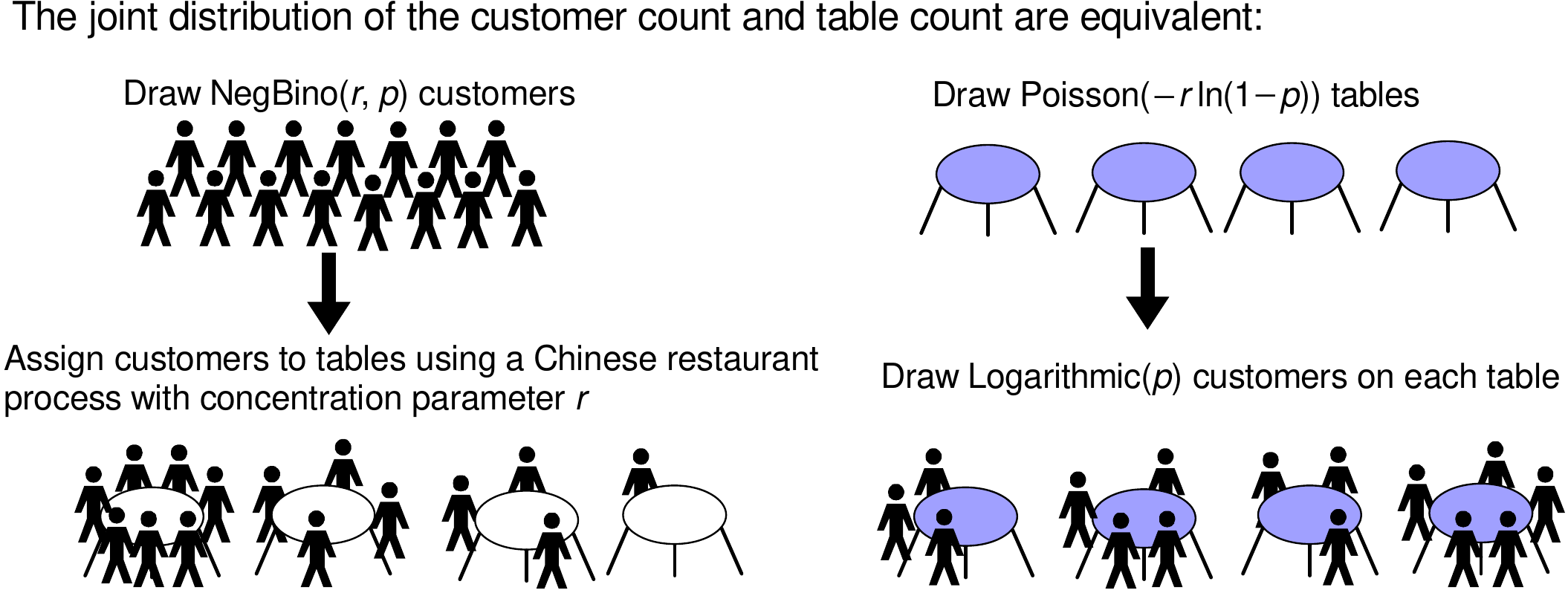}}
\caption{\label{fig:PoisLog}
The Poisson-logarithmic bivariate distribution models the total numbers of customers and tables 
as random variables. As shown in Theorem \ref{lem:PoisLog}, it has two equivalent representations, which connect the Poisson, logarithmic, and negative binomial distributions and the Chinese restaurant process. 
} 
\end{center}
\end{figure}

Advancing previous research on the NB distribution in \cite{LGNB,MingyuanNIPS2012},
    we construct a Poisson-logarithmic bivariate distribution that assists  Bayesian inference of the NB distribution
  and unites various 
   count distributions.
  \begin{thm}[Poisson-logarithmic]\label{lem:PoisLog}
The Poisson-logarithmic (PoisLog) bivariate distribution with PMF
\beqs\label{eq:PMF}
&f_{M,L}(m,l|r,p)
= \frac{|s(m,l)|r^l}{m!}(1-p)^r p^m,
\eeqs
where $m \in\mathbb{Z}_{+}$ and $l = 0,1,\cdots,m$, can be expressed as 
a Chinese restaurant table (CRT) and negative binomial (NB) joint distribution
and also 
a sum-logarithmic and  Poisson joint distribution as
\begin{align}
 \emph{\mbox{PoisLog}}(m,l;r,p)&=  
 \emph{\mbox{CRT}}(l;m,r)\emph{\mbox{NB}}(m;r,p)\notag\\
 &=\emph{\mbox{SumLog}}(m;l,p)\emph{\mbox{Pois}}(l;-r\ln(1-p))\notag,
 \end{align}
 where $\emph{\mbox{SumLog}}(m;l,p)$ denotes the sum-logarithmic distribution generated as
$m = \sum_{{t}=1}^{l}u_{{t}},~ u_{{t}} \stackrel{iid}{\sim} \emph{\mbox{Log}}(p)$.
\end{thm}

  The proof of Theorem \ref{lem:PoisLog} is provided in Appendix A. 
  As shown in Fig.~\ref{fig:PoisLog}, this bivariate distribution 
intuitively describes the
 joint distribution of two count random variables $m$ and $l$ under two equivalent circumstances:
 \begin{itemize}
  \item 1) There are $m\sim\mbox{NB}(r,p)$ customers seated at $l\sim\mbox{CRT}(m,r)$  tables;
 \item 2) There are $l\sim\mbox{Pois}(-r\ln(1-p))$ tables, each of which with $u_t\sim\mbox{Log}(p)$ customers, with $m=\sum_{t=1}^l u_t$ customers in total. 
    \end{itemize}
    In a slight abuse of notation, but for added conciseness, in the following discussion
we use $m\sim \sum_{t=1}^l \mbox{Log}(p)$ to denote $m = \sum_{{t}=1}^{l}u_{{t}},~ u_{{t}} \stackrel{iid}{\sim} \mbox{Log}(p)$.

\begin{cor}\label{lem:merge}  Let $m\sim\emph{\mbox{NB}}(r,p),~r\sim\emph{\mbox{Gamma}}(r_1,1/c_1)$ represent a gamma-NB mixture distribution. It
can be augmented as 
$ 
m\sim \sum_{{t}=1}^{l} \emph{\mbox{Log}}(p),~ l\sim\emph{\mbox{Pois}}(-r\ln(1-p)), ~r\sim\emph{\mbox{Gamma}}(r_1,1/c_1).
$ 
Marginalizing out $r$ leads to 
\beqs\notag
&m\sim\sum_{{t}=1}^{l} \emph{\mbox{Log}}(p),~ l\sim\emph{\mbox{NB}}\left(r_1,p'\right),~p': = \frac{-\ln(1-p)}{c_1-\ln(1-p)},
\eeqs
where the latent count $l\sim\emph{\mbox{NB}}\left(r_1,p'\right)$ 
can be augmented as
\beqs\label{eq:merge}
&
l \sim \sum_{{t}'=1}^{l'}\emph{\mbox{Log}}(p'),~ l' \sim\emph{ \mbox{Pois}}(-r_1\ln(1-p')),\notag
\eeqs
which, using Theorem \ref{lem:PoisLog}, is equivalent in distribution to
\beqs\label{eq:merge1}
 l' \sim\emph{ \mbox{CRT}}(l,r_1),~l \sim \emph{\mbox{NB}}(r_1,p').\notag
\eeqs
\end{cor}

 The connections between various distributions shown in Theorem \ref{lem:PoisLog} and Corollary \ref{lem:merge} are
 key ingredients of this paper, which 
 not only allow us to derive efficient inference, 
 but also, as shown below, let us examine the posteriors to understand fundamental properties of various NB processes,  clearly revealing connections to previous nonparametric Bayesian
mixture models, including those based on the Dirichlet process, HDP and beta-NB processes.

\section{Joint Count and Mixture Modeling}
\label{sec:PP}

In this Section, we first show  the connections between the Poisson and multinomial processes, and then we place a gamma process prior on the Poisson rate measure for joint count and mixture modeling. This construction can be reduced to the Dirichlet process and its restrictions for modeling grouped data are further discussed. 

\subsection{Poisson and Multinomial Processes}\label{sec:4.1}

\begin{cor}\label{cor:PoisMult}
Let $X\sim\emph{\mbox{PP}}(G)$ be a Poisson process defined on a completely random measure $G$ 
such
that $X(A)\sim\emph{\mbox{Pois}}(G(A))$ for each subset $A\subset \Omega$.
Define $Y\sim\emph{\mbox{MP}}(Y(\Omega), \frac{G}{G(\Omega)})$ as a multinomial process, with total count $Y(\Omega)\sim\emph{\mbox{Pois}}(G(\Omega))$  and random probability measure $\frac{G}{G(\Omega)}$, such that $(Y(A_1),\cdots,Y(A_Q))\sim\emph{\mbox{Mult}}\left(Y(\Omega); \frac{G(A_1)}{G(\Omega)},\cdots,\frac{G(A_Q)}{G(\Omega)}\right)$ for any disjoint partition $\{A_q\}_{1,Q}$ of~$\Omega$. 
According to Lemma 4.1 of \emph{\cite{BNBP_PFA}}, $X(A)$ and $Y(A)$ would have the same Poisson distribution for each $A\subset \Omega$, thus $X$ and $Y$ are equivalent in distribution.
\end{cor}

Using Corollary \ref{cor:PoisMult},
we illustrate how the seemingly distinct problems of count
and mixture modeling can be united under the Poisson process.
For each $A\subset \Omega$, denote $X_j(A)$   as a count random variable describing the number of observations in $\xv_j$ that reside within $A$.
Given grouped data $\xv_1,\cdots,\xv_J$, for any measurable disjoint partition  $A_1,\cdots,A_Q$ of $\Omega$, we aim to jointly model  count random variables $\{X_j(A_q)\}$. A natural choice would be to define a Poisson process
\beqs
&X_j\sim\mbox{PP}(G)\notag
\eeqs
with a shared completely random measure $G$ on $\Omega$, such that
$
X_j(A)\sim\mbox{Pois}\big(G(A)\big)
$ for each $A\subset\Omega$ and $G(\Omega) = \sum_{q=1}^Q G(A_q)$. 
 Following Corollary \ref{cor:PoisMult}, with $\widetilde{G} = G/G(\Omega)$, letting $X_j\sim\mbox{PP}(G)$ is equivalent to letting
\beqs\label{eq:Mult}
 &X_j\sim\mbox{MP}(X_j(\Omega), \widetilde{G}),~~X_j(\Omega)\sim\mbox{Pois}(G(\Omega)).\notag
 \eeqs
 Thus the Poisson process provides not only  a way to generate independent counts from each $A_q$, but also a mechanism for mixture modeling, which allocates the $X_j(\Omega)$ observations into any measurable disjoint partition $\{A_q\}_{1,Q}$ of $\Omega$, conditioning on 
 the normalized random measure~$\widetilde{G}$. 

\subsection{Gamma-Poisson Process and Negative Binomial Process}\label{sec:4.2}
To complete the Poisson process, it is natural to place a gamma process prior on the Poisson rate measure $G$ as 
\beqs\label{eq:GaPPP}
\hspace{-4mm}&X_j\sim\mbox{PP}(G),j=1,\cdots,J;~G\sim\mbox{GaP}(J(1-p)/p,G_0).
\eeqs 
For 
a distinct atom $\omega_k$, we have
$n_{jk} \sim\mbox{Pois}(r_k)$,
where $n_{jk}=X_j(\omega_k)$ and $r_k=G(\omega_k)$.
Marginalizing
 out $G$ of the gamma-Poisson process 
leads to an NB process
\beqs
&X=\sum_{j=1}^JX_j\sim\mbox{NBP}(G_0,p) \notag
\eeqs
in which $X(A)\sim\mbox{NB}(G_0(A),p)$ for each $A\subset\Omega$. 

Since $\E[e^{iuX(A)}] = \exp\{{G_0(A)}(\ln(1-p)-\ln(1-pe^{iu}))\} 
= \exp\{G_0(A)\sum_{m=1}^\infty(e^{ium}-1)\frac{p^m}{m}\}$, the L\'{e}vy measure of the NB process can be derived from (\ref{LevyCF}) as
\beqs
&\nu(dn d\omega) = \sum_{m=1}^\infty \frac{p^{m}}{m} \delta_m(dn) G_0(d\omega).\notag
\eeqs
With 
$\nu^{+}=\nu(\mathbb{Z}_{+}\times\Omega)=-\gamma_0\ln(1-p)$, a draw from the NB process consists of a finite number of distinct atoms almost surely and the number of samples on each of them follows a logarithmic distribution, 
expressed as
\beqs\label{eq:GammaDraw}
&X = \sum_{k=1}^{K^+} n_k \delta_{\omega_k},~K^+\sim\mbox{Pois}(-\gamma_0\ln(1-p)),\notag\\
&{(n_k, \omega_k)}\stackrel{iid}{\sim} \mbox{Log}(n_k;p) g_0(\omega_k),~k=1,\cdots,K^+,\label{eq:Log}
\eeqs
where $g_0(d\omega):={G_0(d\omega)}/{\gamma_0}$.
Thus the NB probability parameter $p$ plays a critical role in count and mixture modeling as it directly controls the prior distributions of the number of distinct atoms $K^+\sim\mbox{Pois}(-\gamma_0\ln(1-p))$, the number of samples
at each of these atoms $n_k\sim\mbox{Log}(p)$, and the total number of samples $X(\Omega)\sim\mbox{NB}(\gamma_0,p)$. 
However, its value would become irrelevant if one directly works with the normalization of $G$, as commonly used in conventional mixture modeling.

 Define $L\sim\mbox{CRTP}(X,G_0)$ as a CRT process that
 \beqs\label{eq:Latable}
 &L(A)=\sum_{\omega\in A}L(\omega),~L(\omega)\sim\mbox{CRT}(X(\omega),G_0(\omega))\notag
 \eeqs
  for each $A\subset\Omega$.
  Under the Chinese restaurant process metaphor, $X(A)$ and $L(A)$ represent the customer count and table count, respectively, observed in each $A\subset\Omega$.
  A direct generalization of Theorem \ref{lem:PoisLog} leads to:
   \begin{cor}\label{cor:PoisLog}
 The NB process $X\sim \emph{\mbox{NBP}}(G_0,p)$ augmented under its compound Poisson representation as 
\beqs\label{eq:Aug2_1}
&X \sim \sum_{{t}=1}^{L}\emph{\mbox{Log}}(p), ~L \sim \emph{\mbox{PP}}(-G_0\ln(1-p))\notag
\eeqs
is equivalent in distribution to
\beqs\label{eq:Aug3_1}
&L \sim \emph{\mbox{CRTP}}(X,G_0),~ X\sim \emph{\mbox{NBP}}(G_0,p).\notag 
\eeqs
\end{cor}

\subsection{
Posterior Analysis and Predictive Distribution}
Imposing a gamma prior $\mbox{Gamma}(e_0,1/f_0)$ on $\gamma_0$ 
and a beta prior $\mbox{Beta}(a_0,1/b_0)$ on $p$, using conjugacy, 
we have all conditional posteriors in closed-form as
    \beqs\label{eq:postNB}
    &(G|X,p,G_0) \sim \mbox{GaP}(J/p, G_0 +X)\notag\\
    &(p|X,G_0)\sim\mbox{Beta}(a_0+X(\Omega),b_0+ \gamma_0)\notag\\
    &(L|X,G_0) \sim \mbox{CRTP}(X,G_0)\notag\\
    &(\gamma_0|L,p) \sim\mbox{Gamma}\left(e_0+L(\Omega),\frac{1}{f_0-\ln(1-p)}\right).
    \eeqs
If the  base measure $G_0$ is finite and continuous, then $G_0(\omega)\rightarrow 0$ and we have $L(\omega)\sim\mbox{CRT}(X(\omega),G_0(\omega)) = \delta(X(\omega)>0)$ and thus $L(\Omega) = \sum_{\omega\in\Omega} \delta(X(\omega)>0)$, i.e., the number of nonempty tables $L(\Omega)$  is equal to $K^{+}$, the  number of distinct atoms.
        The gamma-Poisson process is also well defined with a discrete base measure $G_0=\sum_{k=1}^K \frac{\gamma_0}{K} \delta_{\omega_k}$, 
        for which we have  
    $L= \sum_{k=1}^K l_k \delta_{\omega_k},~l_k\sim\mbox{CRT}(X(\omega_k),\gamma_0/K)$ and hence it is possible that $l_k>1$ if $X(\omega_k)>1$, which means $L(\Omega)\ge K^+$. 
    As the data $\{x_{ji}\}_i$ are exchangeable within group~$j$, conditioning on $X^{-ji}=X\backslash{X_{ji}}$ and $G_0$, with $G$ marginalized out, 
     we have
\beqs\label{eq:Predict}
&\hspace{-15mm}X_{ji}|G_0,X^{-ji}\sim \frac{\E[G|G_0,p,X^{-ji}]}{\E[G(\Omega)|G_0,p,X^{-ji}]}\notag \\
&\hspace{20mm}= \frac{G_0}{\gamma_0+X(\Omega)-1}+\frac{X^{-ji}}{\gamma_0+X(\Omega)-1}.
\eeqs
This prediction rule is similar to that of the Chinese restaurant process described in (\ref{eq:CRP}).  

\subsection{Relationship with the Dirichlet Process}
Based on Corollary \ref{cor:PoisMult} on the multinomial process and Section \ref{sec:DP} on the Dirichlet process, the gamma-Poisson process in (\ref{eq:GaPPP}) can be equivalently expressed as 
\beqs\label{eq:PP-DP}
&X_j \sim \mbox{MP}(X_j(\Omega),\widetilde{G}),~\widetilde{G} \sim \mbox{DP}(\gamma_0,  \widetilde{G}_0)\notag\\
&X_j(\Omega)\sim\mbox{Pois}(\alpha),~\alpha\sim\mbox{Gamma}(\gamma_0,p/(J(1-p))),
\eeqs
where $G=\alpha \widetilde{G}$ and $G_0 = \gamma_0\widetilde{G}_0$. Thus without modeling $X_j(\Omega)$ and $\alpha=G(\Omega)$ as random variables,
the gamma-Poisson process becomes a Dirichlet process, which is widely used for mixture modeling \cite{ferguson73,Escobar1995,MullerDP,ishwaran02,Teh2010a}.
 Note that for the Dirichlet process, the inference of $\gamma_0$ relies on a data augmentation method of 
  \cite{Escobar1995} when $\widetilde{G}_0$ is continuous, and no rigorous inference for $\gamma_0$ is available when $\widetilde{G}_0$ is discrete.
Whereas for  the proposed gamma-Poisson process augmented from the NB process, 
as shown in (\ref{eq:postNB}), regardless of whether the base measure $G_0$ is continuous or discrete, $\gamma_0$ has an analytic conditional gamma posterior, 
with conditional expectation 
$\E[\gamma_0|L,p] = \frac{e_0+L(\Omega)}{f_0-\ln(1-p)}. 
$ 

\subsection{Restrictions of the Gamma-Poisson Process}
 The Poisson process has an equal-dispersion assumption for count modeling. 
 For mixture modeling of grouped data, the gamma-Poisson (NB) process 
 might be too restrictive in  that, 
 as shown in (\ref{eq:PP-DP}),  it implies the same mixture proportions across groups, and as shown in (\ref{eq:Log}),  it implies the same 
 count distribution on each distinct atom. 
 This motivates us to consider adding an additional layer into the gamma-Poisson process
 or using a different distribution other than the Poisson to model the counts for grouped data. 
 As shown in Section \ref{sec:CountDist}, the NB distribution is an ideal candidate, not only because it allows overdispersion, 
 but also because it can be augmented into either a gamma-Poisson or a compound Poisson representations and it can be used together with the CRT distribution to form a  bivariate distribution that jointly models the counts of customers and tables.


%

\section{Joint Count and Mixture Modeling of Grouped Data} 
\label{sec:GNBP} 

In this Section we couple the gamma process with the NB process to construct a gamma-NB process, which is well suited for modeling grouped data. We derive analytic conditional posteriors for this construction and show that it can be reduced to an HDP. 

\subsection{Gamma-Negative Binomial Process}

For joint count and mixture modeling of grouped data, e.g., topic modeling where a document consists of a group of exchangeable words, we replace the Poisson processes in (\ref{eq:GaPPP}) with NB processes. Sharing the NB dispersion across groups while making the probability parameters be group dependent,
we  construct a gamma-NB process
as
\beqs\label{eq:GammaNP}
&X_j \sim \mbox{NBP}(G,p_j),~ G\sim \mbox{GaP}(c, G_0).
\eeqs
With $G\sim \mbox{GaP}(c, G_0)$
expressed as 
$G=\sum_{k=1}^\infty r_k \delta_{\omega_k}$, a draw from 
$\mbox{NBP}(G,p_j)$ 
can be expressed as 
$X_j = \sum_{k=1}^\infty n_{jk}\delta_{\omega_k}, ~n_{jk}\sim \mbox{NB}(r_k,p_j).$

\begin{figure}[!tb]
\vskip 0.077in
\begin{center}
\includegraphics[width=0.98\columnwidth]
{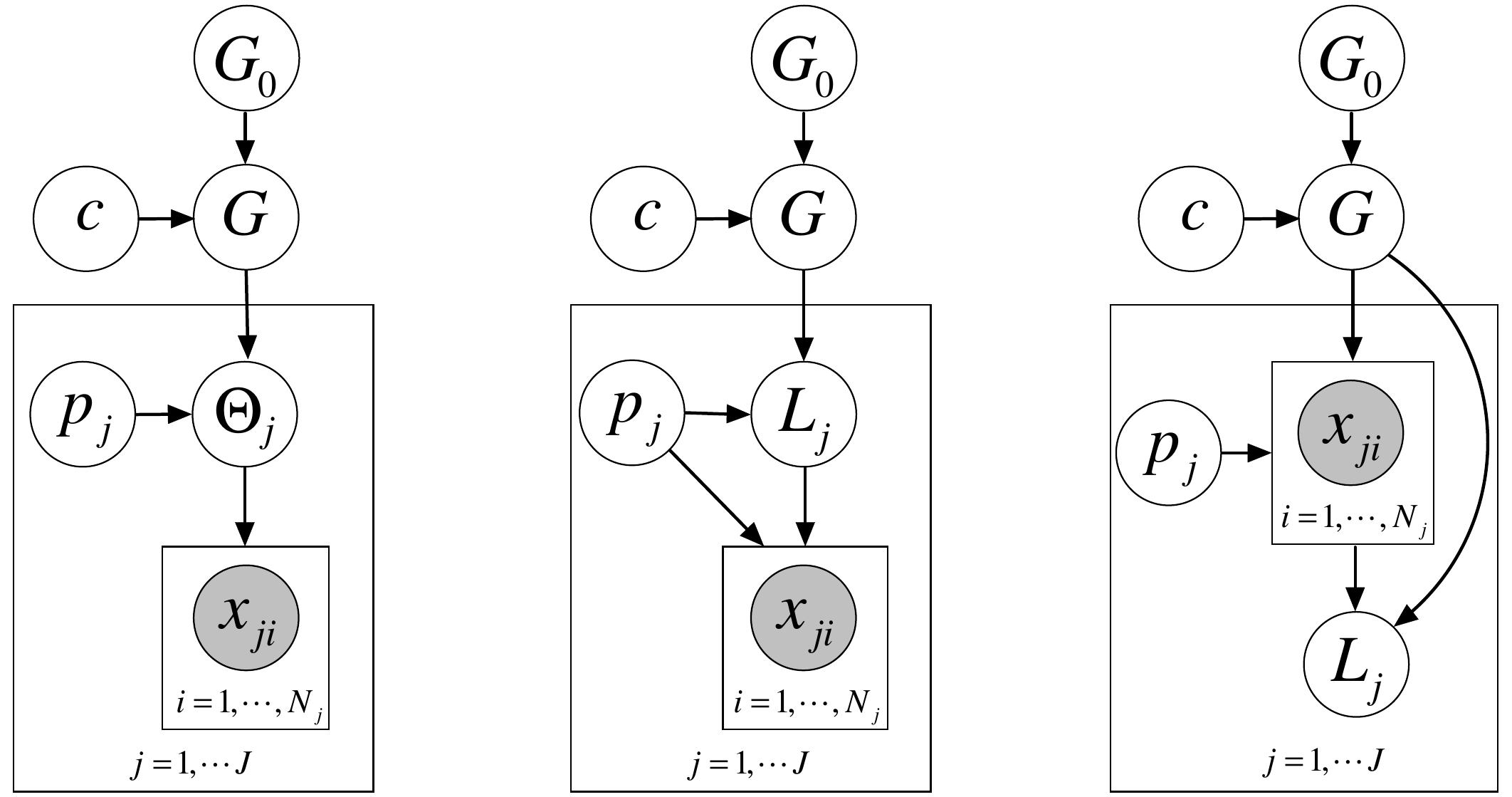}
\end{center}
\caption{ \label{fig:GammaNBP}
Graphical models of the gamma-negative binomial process under the gamma-gamma-Poisson (left), gamma-compound Poisson (center), and gamma-negative binomial-Chinese restaurant table constructions (right). The center and right constructions are equivalent in distribution.
}
\end{figure}

The gamma-NB process can be augmented as a gamma-gamma-Poisson process as
\beqs\label{eq:HGP-PP}\small
\hspace{-4mm}&X_j \sim \mbox{PP}(\Theta_j),~\Theta_j \sim \mbox{GaP}\big(\frac{1-p_j}{p_j}, G\big), ~G\sim \mbox{GaP}(c, G_0)
\eeqs
and with $\theta_{jk}=\Theta_j(\omega_k)$, 
we have
$
n_{jk}\sim\mbox{Pois}(\theta_{jk}),~\theta_{jk}\sim\mbox{Gamma}(r_k,p_j/(1-p_j)).
$
 This construction introduces gamma processes $\{\Theta_j\}$, whose normalization provide group-specific random probability measures $\{\widetilde{\Theta}_j\}$ for mixture modeling.
  The gamma-NB process can also be augmented as 
\beqs
&\label{eq:Aug2} X_j \sim \sum_{{t}=1}^{L_j}\mbox{Log}(p_j), ~L_j \sim \mbox{PP}(-G\ln(1-p_j)),\notag\\
&G\sim \mbox{GaP}(c, G_0),
\eeqs
which is equivalent in distribution to
\beqs\label{eq:Aug3}
\hspace{-6mm}&L_j \sim \mbox{CRTP}(X_j,G),X_j\hspace{-0.5mm}\sim \hspace{-.5mm}\mbox{NBP}(G,p_j),G\sim \mbox{GaP}(c, G_0)
\eeqs
according to Corollary \ref{cor:PoisLog}.
These three closely related constructions are graphically presented in Fig. \ref{fig:GammaNBP}.

With Corollaries \ref{lem:merge} and \ref{cor:PoisLog}, $p': = \frac{-\sum_j\ln(1-p_j)}{c-\sum_j\ln(1-p_j)}$ and $L: = \sum_j L_j$, we further have two equivalent augmentations:
\beqs
& \label{eq:Laugment} L\sim\sum_{{t}=1}^{L'}\mbox{Log}(p'), ~ L' \sim \mbox{PP}(-G_0\ln(1-p'));\\
&\label{eq:Aug4}  L' \sim \mbox{CRTP}(L,G_0),~ L\sim \mbox{NBP}(G_0,p').
\eeqs
These augmentations allow us to derive a sequence of  closed-form update equations, 
as described below.

\subsection{Posterior Analysis and Predictive Distribution}\label{sec:post}

With $p_j\sim\mbox{Beta}(a_0,b_0)$ and (\ref{eq:GammaNP}), 
we have
\beq
(p_j|-)\sim \mbox{Beta}\left(a_0+X_j(\Omega),b_0+ G(\Omega)\right).
\eeq
Using (\ref{eq:Aug3}) and (\ref{eq:Aug4}), 
we have
\beqs
&L_j|X_j,G \sim \mbox{CRTP}(X_j,G),\\
&L'|L,G_0 \sim \mbox{CRTP}(L,G_0).
\eeqs
If $G_0$ is finite and continuous, we have $G_0(\omega)\rightarrow 0$ $\forall~\omega\in \Omega$  and thus
$
L'(\Omega)|L,G_0 = \sum_{\omega\in\Omega}\delta(L(\omega)>0)= \sum_{\omega\in\Omega}\delta( \sum_{j}X_j(\omega)>0)=K^+;
$
if $G_0$ is discrete as $G_0=\sum_{k=1}^K\frac{\gamma_0}{K}\delta_{\omega_k}$, 
then $L'(\omega_k)=\mbox{CRT}(L(\omega_k),\frac{\gamma_0}{K})\ge 1$ if $\sum_{j}X_j(\omega_k)>0$, thus $L'(\Omega)\ge K^{+}$. In either case,
let $\gamma_0=G_0(\Omega) \sim{\mbox{Gamma}}(e_0,1/f_0)$, 
with  the gamma-Poisson conjugacy on (\ref{eq:Laugment}) and (\ref{eq:Aug2}), 
we have
\beqs
\hspace{-4mm}&\label{eq:gamma0post}\gamma_0|\{L'(\Omega),p'\} \sim {\mbox{Gamma}}\left( e_0 +  L'(\Omega),\frac{1}{f_0- \ln(1-p')}\right),\\
\label{eq:Gpost}
\hspace{-4mm}&G|G_0,\{L_j,p_j\} \sim {\mbox{GaP}} \left(c - \sum_{j} \ln( 1 - p_j), G_0 +  L \right).
\eeqs
Using the  gamma-Poisson conjugacy on (\ref{eq:HGP-PP}), 
we have
\beq
\Theta_j|G,X_j,p_j \sim {\mbox{GaP}}\left(1/p_j,G+X_j\right).
\eeq

Since the data $\{x_{ji}\}_i$ are exchangeable within group~$j$, conditioning on $X_j^{-i}=X_j\backslash X_{ji}$ 
and $G$, with $\Theta_j$ marginalized out, we have 
\beqs\label{eq:GammaNBPolyaUrn}
&\hspace{-24mm}X_{ji}|G,X_j^{-i}\sim \frac{\E[\Theta_j|G,X_j^{-i}]}{\E[\Theta_j(\Omega)|G,X_j^{-i}]} \notag\\
&\hspace{20mm}= \frac{G}{G(\Omega)+X_j(\Omega)-1}+\frac{X_j^{-i}}{G(\Omega)+X_j(\Omega)-1}.
\eeqs
This prediction rule is similar to that of the  Chinese restaurant franchise (CRF) \cite{HDP}. 

\subsection{Relationship with Hierarchical Dirichlet Process}

With Corollary \ref{cor:PoisMult} and Section \ref{sec:DP}, we can equivalently express the gamma-gamma-Poisson process in (\ref{eq:HGP-PP}) 
as 
\beqs
\label{eq:PP-HDP1}
&X_j \sim \mbox{MP}(X_j(\Omega),\widetilde{\Theta}_j),~\widetilde{\Theta}_j \sim \mbox{DP}(\alpha,  \widetilde{G}),\notag\\
\label{eq:PP-HDP}
&X_j(\Omega)\sim\mbox{Pois}(\theta_j),~\theta_j\sim\mbox{Gamma}(\alpha,p_j/(1-p_j)),\notag\\
&\alpha \sim\mbox{Gamma}(\gamma_0,1/c),~\widetilde{G}\sim \mbox{DP}(\gamma_0, \widetilde{G}_0),
\eeqs
where $\Theta_j=\theta_j \widetilde{\Theta}_j$, $G=\alpha \widetilde{G}$ and $G_0 = \gamma_0\widetilde{G}_0$.
Without modeling $X_j(\Omega)$ and $\theta_j$ as random variables, (\ref{eq:PP-HDP}) becomes an HDP \cite{HDP}.
Thus the augmented and then normalized gamma-NB process leads to an HDP.
However, we cannot return from the HDP to the gamma-NB process without modeling $X_j(\Omega)$ 
and $\theta_j$ as 
random variables.
Theoretically, they are distinct in that the gamma-NB process is a completely random measure, assigning independent random variables into any disjoint Borel sets $\{A_q\}_{1,Q}$ in $\Omega$, and the count $X_j(A)$ has the distribution as $X_j(A)\sim\mbox{NB}(G(A),p_j)$; by contrast, due to normalization, the HDP 
is not, and marginally
\beq
X_j(A)\sim\mbox{Beta-Binomial}\big(X_j(\Omega),\alpha\widetilde{G}(A),\alpha(1-\widetilde{G}(A))\big).\notag
\eeq
Practically, the gamma-NB process can exploit Corollary \ref{cor:PoisLog} and the gamma-Poisson conjugacy to achieve analytic conditional posteriors. 
The inference of the HDP is a  challenge and it is usually solved through alternative constructions such as the CRF and stick-breaking representations \cite{HDP,VBHDP}.
In particular, 
both  concentration parameters $\alpha$ and $\gamma_0$ are nontrivial  to infer \cite{HDP,HDP-HMM} and they are often simply fixed  \cite{VBHDP}. One may apply the data augmentation method of \cite{Escobar1995} to sample $\alpha$ and $\gamma_0$. However, if $\widetilde{G}_0$ is discrete as $\widetilde{G}_0=\sum_{k=1}^K \frac{1}{K}\delta_{\omega_k}$, which is of practical value and becomes a continuous base measure as $K\rightarrow \infty$ \cite{ishwaran02,HDP,HDP-HMM}, then using that method 
to sample $\gamma_0$ is only approximately correct, which may result in a biased estimate in practice, especially if $K$ is not sufficiently large. 

By contrast, in the gamma-NB process, the shared 
 $G$ can be analytically updated with (\ref{eq:Gpost})
and $G(\Omega)$ plays the role of $\alpha$ in the HDP, which is readily available as
\beqs \label{eq:alphaPost}
&(G(\Omega)|-)
\sim \mbox{Gamma}  {\left( \gamma_0 +   L(\Omega), \frac{1}{c - \sum_j \ln( 1 - p_j)}\right)}
\eeqs
and as in (\ref{eq:gamma0post}), regardless of whether the base measure is continuous, the total mass $\gamma_0$ has an analytic gamma posterior. Equation (\ref{eq:alphaPost}) also intuitively shows how the NB probability parameters $\{p_j\}$ govern the variations among $\{\widetilde{\Theta}_j\}$ in the gamma-NB process.
In the HDP, $p_j$ is not explicitly modeled, and since its value appears irrelevant when taking the normalized constructions in (\ref{eq:PP-HDP}), it is usually  treated as a nuisance parameter  and perceived as $p_j=0.5$ when needed
for interpretation. 

Another related model is the DILN-HDP 
in \cite{DILN}, where group-specific Dirichlet processes are normalized from gamma processes, with the gamma scale parameters either fixed as $\frac{p_j}{1-p_j}=1$ or learned with a log Gaussian process prior. Yet no analytic conditional posteriors are provided and Gibbs sampling is not considered as a viable option. The main purpose of \cite{DILN} is introducing correlations between mixture components. 
It would be interesting to compare the differences between 
learning the $\{p_j\}$ with beta priors 
and learning the gamma scale parameters with the log Gaussian process prior. 

\section{The Negative Binomial Process Family}\label{sec:BNB}

The gamma-NB process 
shares the NB dispersion across groups while the NB probability parameters are group dependent. Since the NB distribution has two adjustable parameters, it is natural to wonder whether one can explore sharing the NB probability measure across groups, while making the NB dispersion parameters group specific or atom dependent.
That kind of construction would be distinct from both the gamma-NB process and HDP in that $\Theta_j$ has space dependent scales, and thus its normalization $\widetilde{\Theta}_j= \frac{\Theta_j}{\Theta_j(\Omega)}$, still a random probability measure, no longer follows a Dirichlet process.

 It is natural to let the NB probability measure be drawn from the beta process \cite{Hjort,JordanBP}. In fact, the first discovered member of the NB process family is a beta-NB process \cite{BNBP_PFA}. A main motivation of that construction
 is observing that the beta and Bernoulli distributions are conjugate and the beta-Bernoulli process is found to be quite useful for dictionary learning 
 \cite{MingyuanSAM2010,dHBP,DictTopic,BPFA2012}, whereas although the beta distribution is also conjugate to the NB distribution, there is apparent lack of exploitation of that relationship  \cite{BNBP_PFA}. 
 
A beta-NB process \cite{BNBP_PFA,NBPJordan} is constructed by letting
\beqs
&X_j\sim\mbox{NBP}(r_j,B),~B\sim\mbox{BP}(c,B_0).
\eeqs
With $B\sim\mbox{BP}(c,B_0)$ expressed as $B=\sum_{k=1}^\infty p_k\delta_{\omega_k}$, a random draw from $\mbox{NBP}(r_j,B)$ can be
expressed as
\beqs
&X_j = \sum_{k=1}^\infty n_{jk}\delta_{\omega_k}, ~n_{jk}\sim \mbox{NB}(r_j,p_k).
 \eeqs
 Under this construction, the NB probability measure is shared  and the NB dispersion parameters are group dependent. Note that if $\{r_j\}$ are fixed as one, then the beta-NB process reduces to the beta-geometric process, related to the one for count modeling discussed in \cite{Thibaux}; if $\{r_j\}$  are empirically set to some other values, 
then the beta-NB process reduces to the one proposed in \cite{NBPJordan}.  These simplifications are not considered in the paper, as they are often overly restrictive.

The asymptotic behavior of the beta-NB process with respect to the NB dispersion parameter is studied in \cite{NBPJordan}. Such analysis is not provided here  as 
we infer NB dispersion parameters from the data, which usually do not have large values due to overdispersion. 
In \cite{NBPJordan}, the beta-NB process is treated comparable to a gamma-Poisson process and is considered less flexible than the HDP, motivating the construction of a hierarchical-beta-NB process. By contrast, in this paper, with the beta-NB process augmented as a beta-gamma-Poisson process, one can draw group-specific Poisson rate measures for count modeling and then use their normalization 
to provide group-specific random probability measures for mixture modeling; therefore, the beta-NB process, gamma-NB process and HDP are treated comparable to each other in hierarchical structures and are all considered suitable for mixed-membership modeling. 

As in \cite{BNBP_PFA}, we may also consider a marked-beta-NB process, with both the NB probability and dispersion measures shared, in which each point of the beta process is marked with an independent gamma random variable. Thus a draw from the marked-beta process becomes $(R,B) = \sum_{k=1}^\infty (r_k,p_k) \delta_{\omega_k}$, and a draw from the NB process $X_j\sim \mbox{NBP}(R,B)$ becomes
\beqs
&X_j = \sum_{k=1}^\infty n_{jk}\delta_{\omega_k}, ~n_{jk}\sim \mbox{NB}(r_k,p_k).
\eeqs
With the beta-NB conjugacy, the posterior of $B$ is tractable in both the beta-NB and marked-beta-NB processes \cite{BNBP_PFA,NBPJordan,MingyuanNIPS2012}. 
Similar to the marked-beta-NB process, we may also consider a marked-gamma-NB process, where each point of the gamma process is marked with an independent beta random variable, whose performances is found to be similar.

If it is believed that there are  excessive number of zeros, governed by  a  process other than the NB process, we may introduce a zero inflated NB process as $X_j \sim \mbox{NBP}(RZ_j,p_j)$, where $Z_j\sim \mbox{BeP}(B)$ is drawn from the Bernoulli process \cite{JordanBP} and $(R,B) = \sum_{k=1}^\infty (r_k,\pi_k)\delta_{\omega_k}$ is drawn from a gamma marked-beta process, thus  a draw from $\mbox{NBP}(RZ_j,p_j)$ can be expressed as $X_j = \sum_{k=1}^\infty n_{jk}\delta_{\omega_k}$, with
\beqs
&n_{jk}\sim \mbox{NB}(r_k b_{jk}, p_j), ~b_{jk}  = \mbox{Bernoulli}(\pi_k).
\eeqs
 This construction can be linked to the focused topic model in \cite{FocusTopic} with appropriate normalization, with advantages that there is no need to fix $p_j=0.5$ and the inference is fully tractable. The zero inflated construction can also be linked to models for real valued data using the Indian buffet process (IBP) or beta-Bernoulli process spike-and-slab prior \cite{IBP,knowles,Paisley_ICML,Mingyuan09}. 
 Below we apply various NB processes for topic modeling and illustrate the differences between them.

\section{Negative Binomial Process Topic Modeling and Poisson Factor Analysis} \label{sec:PFA}

We consider topic modeling of a document corpus, a special case of mixture modeling of grouped data, where the words of the $j$th document $x_{j1},\cdots,x_{jN_j}$ constitute a group $\xv_j$  ($N_j$ words in document $j$), each word $x_{ji}$ is an exchangeable group member indexed by $v_{ji}$ in a vocabulary with $V$ unique terms. Each word $x_{ji}$ is drawn from a topic $\phiv_{z_{ji}}$ as $x_{ji}\sim F(\phiv_{z_{ji}})$, where $z_{ji}=1,2,\cdots,\infty$ is the topic index and the likelihood $F(x_{ji};\phiv_k)$ is simply $\phi_{v_{ji}k}$, the probability of word $x_{ji}$ under topic $\phiv_k = (\phi_{1k},\cdots,\phi_{Vk})^T \in \mathbb{R}_{+}^{V}$, with $\sum_{v=1}^V{\phi_{vk}}=1$.  We refer to NB process mixture modeling of grouped words $\{\xv_j\}_{1,J}$ as NB process topic modeling.

For the gamma-NB process described in Section \ref{sec:GNBP}, 
with the gamma process expressed as $G=\sum_{k=1}^\infty r_k\delta_{\phiv_k}$,
we can express the hierarchical model as
\beqs
&x_{ji}\sim F(\phiv_{z_{ji}}),~\phiv_k\sim g_0(\phiv_k),~N_j = \sum_{k=1}^{\infty} n_{jk},\notag\\
&\hspace{-0mm}n_{jk}\sim \mbox{Pois}(\theta_{jk}), ~\theta_{jk}\sim \mbox{Gamma}(r_k,p_j/(1-p_j))
\label{eq:GNBP_mixture0}
\eeqs
where $g_0(d\phiv)= G_0(d\phiv)/\gamma_0$.
With $\theta_{j} = \Theta_j(\Omega) = \sum_{k=1}^\infty\theta_{jk}$, $\nv_j = (n_{j1},\cdots,n_{j\infty})^T$ and $\thetav_j = (\theta_{j1},\cdots,\theta_{j\infty})^T$, using Corollary \ref{cor:PoisMult}, we can equivalently express $N_j$ and $n_{jk}$
in (\ref{eq:GNBP_mixture0}) as
\beqs
N_j\sim \mbox{Pois}\left(\theta_{j} \right),~ \nv_j \sim \mbox{Mult}\left(N_j;\thetav_j/\theta_{j}\right). \label{eq:nvj}
\eeqs
Since $\{x_{ji}\}_{i=1,N_j}$  are fully exchangeable, rather than drawing $\nv_j$ as in (\ref{eq:nvj}), we may equivalently draw it as 
\beqs \label{eq:z_ji}
\hspace{-5mm}&z_{ji} \sim \mbox{Discrete}(\thetav_j/\theta_{j})
,~n_{jk} = \sum_{i=1}^{N_j}\delta(z_{ji}=k).
\eeqs
This provides further insights on uniting 
the seemingly distinct problems of count and mixture modeling. 

Denote $n_{vjk}=\sum_{i=1}^{N_j}\delta(z_{ji}=k,v_{ji}=v)$, 
$n_{v\cdotv k}=\sum_{j}n_{vjk}$ and $n_{\cdotv k}=\sum_{j}n_{jk}$.
For modeling convenience, we place Dirichlet priors on topics $\phiv_k\sim \mbox{Dir}(\eta,\cdots,\eta)$, then for the gamma-NB process topic model, we have
\beqs\label{eq:inference_z}
&\mbox{Pr}(z_{ji}=k|-) \propto 
\phi_{v_{ji}k}\theta_{jk},\\ 
\label{eq:inference_phi}
&(\phiv_k|-) \sim \mbox{Dir}\left(\eta + n_{1\cdotv k},\cdots,\eta+ n_{V\cdotv k}\right),
\eeqs
which would be the same for the other NB processes, since the gamma-NB process differs from them only on how the gamma priors of $\theta_{jk}$ and consequently the NB priors of $n_{jk}$ are constituted. For example, marginalizing out $\theta_{jk}$, we have $n_{jk}\sim\mbox{NB}(r_k,p_j)$ for the gamma-NB process, $n_{jk}\sim\mbox{NB}(r_j,p_k)$ for the beta-NB process, $n_{jk}\sim\mbox{NB}(r_k,p_k)$ for both the marked-beta-NB and marked-gamma-NB processes, and $n_{jk}\sim \mbox{NB}(r_k b_{jk}, p_j)$ for the zero-inflated-NB process.

\subsection{Poisson Factor Analysis} 

Under the bag-of-words representation, 
without losing information, we can form $\{\xv_j\}_{1,J}$ as a term-document count matrix $\Mmat\in \mathbb{R}^{V\times J}$, where $m_{vj}$ counts the number of times term $v$ appears in document $j$.
Given $K\le \infty$ and 
$\Mmat$, discrete latent variable models assume that each entry $m_{vj}$ can be explained as a sum of smaller counts, each produced by one of the $K$ hidden factors, or in the case of topic modeling, a hidden topic. We can factorize $\Mmat$ under the Poisson likelihood as
\beq\label{eq:pfa}
\Mmat \sim \mbox{Pois}(\Phimat \Thetamat),\notag
\eeq
where $\Phimat\in\mathbb{R}^{V\times K}$ is the factor loading matrix, each column of which is an atom encoding the relative importance of each term, and $\Thetamat\in\mathbb{R}^{K\times J}$ is the factor score matrix, each column of which encodes the relative importance of each atom in a sample. 
This is called Poisson Factor Analysis~(PFA) \cite{BNBP_PFA}.

As in \cite{Dunson05bayesianlatent,BNBP_PFA}, we may augment $m_{vj}\sim\mbox{Pois}(\sum_{k=1}^K\phi_{vk}\theta_{jk})$ as
\beqs\label{eq:PoisAug}
&m_{vj} = \sum_{k=1}^K n_{vjk}, ~n_{vjk} \sim \mbox{Pois}(\phi_{vk}\theta_{jk}).\notag
\eeqs
If  $\sum_{v=1}^V{\phi_{vk}}=1$, we have $n_{jk}\sim\mbox{Pois}(\theta_{jk})$, and with Corollary \ref{cor:PoisMult} 
and $\thetav_j=(\theta_{j1},\cdots,\theta_{jK})^T$, we  also have
$(n_{vj1},\cdots,n_{vjK}|-)\sim \mbox{Mult}\big(m_{vj}; \frac{\phi_{v1}\theta_{j1}}{\sum_{k=1}^K\phi_{vk}\theta_{jk}},$ $\cdots,\frac{\phi_{vK}\theta_{jK}}{\sum_{k=1}^K\phi_{vk}\theta_{jk}}  \big)$, $(n_{v\cdotv 1},\cdots,n_{v\cdotv K}|-)$ $\sim \mbox{Mult}(n_{\cdotv k};\phiv_{k})$, and  $(n_{j1},\cdots,n_{jK}|-)$ $\sim \mbox{Mult}\left(N_j; \thetav_j\right)$, which would lead to (\ref{eq:inference_z}) under the assumption that the words $\{x_{ji}\}_i$ are exchangeable and (\ref{eq:inference_phi}) if $\phiv_k\sim \mbox{Dir}(\eta,\cdots,\eta)$.
Thus topic modeling with the NB process  can be considered as factorization of the term-document count matrix under the Poisson likelihood as $\Mmat\sim\mbox{Pois}(\Phimat\Thetamat)$. 

PFA provides a unified framework to connect previously proposed discrete latent variable models, such as those in \cite{LDA,NMF,CannyGaP,DCA,FocusTopic}. As discussed in detail in \cite{BNBP_PFA}, 
these models mainly differ on how the priors of $\phi_{vk}$ and $\theta_{jk}$ are constituted and how the inferences are implemented. For example, nonnegative matrix factorization \cite{NMF} with an objective function of minimizing the Kullback-Leibler (KL) divergence $D_\text{KL}(\Mmat||\Phimat\Thetamat)$ is equivalent to the ML estimation of $\Phimat$ and $\Thetamat$ under PFA, and latent Dirichlet allocation (LDA) \cite{LDA} is equivalent to a PFA with Dirichlet priors imposed on both $\phiv_k$ and $\thetav_j$. 

\subsection{Negative Binomial Process Topic Modeling}
\begin{table*}\caption{A variety of negative binomial processes are constructed with distinct sharing mechanisms, reflected with which parameters from 
$r_k$, $r_j$, $p_k$, $p_j$ and $\pi_k$ ($b_{jk}$) are inferred (indicated by a check-mark~$\checkmark$), and the implied variance-mean-ratio (VMR) and overdispersion level (ODL)
for counts $\{n_{jk}\}_{j,k}$. They are applied for topic modeling, 
a typical example of mixture modeling of grouped data. Related  algorithms are shown in the last column.}
\centering
{
\begin{tabular}{|c|c|c|c|c|c|c|c|c|c|}
  \hline
  Algorithms                   &$\theta_{jk}$  & $r_k$ & $r_j$ & $p_k$ & $p_j$ & $\pi_k$ & VMR & ODL & Related Algorithms\\ \hline
  NB     &  $\theta_{jk} \equiv r_k$     & $\checkmark$ &  &  &  & & $(1-p)^{-1}$ & $r_k^{-1}$ & Gamma-Poisson \cite{InfGaP}, Gamma-Poisson \cite{Thibaux} \\\hline
  NB-LDA            & $\checkmark$ &  & $\checkmark$ &  & $\checkmark$ & & $(1-p_j)^{-1}$ & $r_j^{-1}$ &LDA \cite{LDA}, Dir-PFA \cite{BNBP_PFA} \\\hline
  NB-HDP            & $\checkmark$       & $\checkmark$ & &  & $0.5$ & & $2$ & $r_k^{-1}$ &HDP \cite{HDP}, DILN-HDP \cite{DILN} \\ \hline
  NB-FTM       & $\checkmark$& $\checkmark$ &  & & $0.5$ & $\checkmark$ & $2$ & $(r_k)^{-1}b_{jk}$& FTM \cite{FocusTopic}, S$\gamma\Gamma$-PFA \cite{BNBP_PFA}\\  \hline
 Beta-Geometric    & $\checkmark$    &  & 1  & $\checkmark$ &  & & $(1-p_k)^{-1}$ & 1 &  Beta-Geometric \cite{Thibaux}, BNBP \cite{BNBP_PFA}, BNBP \cite{NBPJordan}\\ \hline
  Beta-NB       & $\checkmark$ &  & $\checkmark$  & $\checkmark$ &  & & $(1-p_k)^{-1}$ & $r_j^{-1}$ &  BNBP \cite{BNBP_PFA}, BNBP \cite{NBPJordan}\\ \hline
  Gamma-NB       & $\checkmark$            & $\checkmark$ & &  & $\checkmark$ & & $(1-p_j)^{-1}$ & $r_k^{-1}$ & CRF-HDP \cite{HDP,HDP-HMM} \\ \hline
  Marked-Beta-NB    & $\checkmark$   & $\checkmark$ &  & $\checkmark$ &  & & $(1-p_k)^{-1}$ & $r_k^{-1}$ & BNBP \cite{BNBP_PFA}   \\
  \hline
\end{tabular}\label{Tab:Relationships}
}
\end{table*}

%

From the point view of PFA, an NB process topic model factorizes the term-document count matrix under the constraints that each factor sums to one and the factor scores are gamma distributed random variables, and consequently, the number of words assigned to a topic (factor/atom) follows an NB distribution. Depending on how
the NB distributions are parameterized, as shown in Table \ref{Tab:Relationships}, we can construct a variety of NB process topic models, which can also be connected to a large number of previously proposed  parametric and nonparametric topic models. For a deeper understanding on how the counts are modeled, we also show in Table \ref{Tab:Relationships} both the variance-to-mean ratio (VMR) and overdispersion level (ODL) implied by these settings.
Eight differently constructed NB processes are considered: 
 \begin{itemize}
\item

(\emph{i}) The NB process described in Section \ref{sec:PP}
is used for topic modeling. 
It improves over the count-modeling gamma-Poisson process discussed in \cite{InfGaP,Thibaux} in that it unites mixture modeling and has closed-form conditional posteriors. Although this is a nonparametric model supporting an infinite number of  topics, requiring $\{\theta_{jk}\}_{j=1,J}\equiv r_k$ may be too restrictive.

\item
(\emph{ii}) Related to LDA \cite{LDA} 
and Dir-PFA \cite{BNBP_PFA}, the NB-LDA is also a parametric topic model that requires tuning the number of topics. It is constructed by replacing the topic weights of the Gamma-NB process in (\ref{eq:GNBP_mixture0}) as $\theta_{jk}\sim\mbox{Gamma}(r_j,p_j/(1-p_j))$.  It uses document dependent $r_j$ and $p_j$ to 
learn the smoothing of the topic weights, 
and it lets $r_j\sim\mbox{Gamma}(\gamma_0,1/c),~\gamma_0\sim\mbox{Gamma}(e_0,1/f_0)$ to share statistical strength between documents. 
\item

(\emph{iii}) Related to the HDP \cite{HDP}, the NB-HDP model is constructed by fixing $p_j/(1-p_j)\equiv 1$ (i.e., $p_j\equiv 0.5$) in (\ref{eq:GNBP_mixture0}). 
It is also comparable to the HDP in \cite{DILN} that constructs group-specific Dirichlet processes with normalized gamma processes, whose scale parameters are also set as one. 
\item

(\emph{iv}) The NB-FTM model is constructed by replacing the topic weights in (\ref{eq:GNBP_mixture0}) as $\theta_{jk}\sim\mbox{Gamma}(r_kb_{jk},p_j/(1-p_j))$, with $p_j\equiv 0.5$ and $b_{jk}$ drawn from a beta-Bernoulli process 
 that is used to explicitly model zero counts.
It is the same as the sparse-gamma-gamma-PFA (S$\gamma\Gamma$-PFA) in \cite{BNBP_PFA} and is comparable to the focused topic model (FTM) \cite{FocusTopic}, which is constructed from the IBP compound Dirichlet process. 
The Zero-Inflated-NB process improves over these approaches by allowing $\{p_j\}$ to be inferred, which generally yields better data fitting.

(\emph{v}) The Gamma-NB process, as shown in (\ref{eq:GammaNP}) and (\ref{eq:GNBP_mixture0}),  explores sharing the NB  dispersion measure across groups, and it improves over the NB-HDP by allowing the learning of $\{p_j\}$. As shown in (\ref{eq:PP-HDP}), it reduces to the  HDP in \cite{HDP}  without modeling $X_j(\Omega)$ and $\theta_j$ as random variables.
\item

(\emph{vi}) The Beta-Geometric process is constructed by replacing the topic weights in (\ref{eq:GNBP_mixture0}) as $\theta_{jk}\sim\mbox{Gamma}(1,p_k/(1-p_k))$. It explores sharing the NB probability measure across groups, which is related to the one proposed for count modeling in \cite{Thibaux}. It is restrictive that the NB dispersion parameters are fixed as one.
\item

(\emph{vii}) The Beta-NB process is constructed by replacing the topic weights in (\ref{eq:GNBP_mixture0}) as $\theta_{jk}\sim\mbox{Gamma}(r_j,p_k/(1-p_k))$. It explores sharing the NB probability measure across groups, which improves over the Beta-Geometric process and the beta-NB process (BNBP) proposed in \cite{NBPJordan} by providing analytic conditional posteriors of~$\{r_j\}$.

\item
(\emph{viii}) The Marked-Beta-NB process constructed by replacing the topic weights in (\ref{eq:GNBP_mixture0}) as $\theta_{jk}\sim\mbox{Gamma}(r_k,p_k/(1-p_k))$. It is comparable to the BNBP proposed in \cite{BNBP_PFA}, with the distinction that it provides analytic conditional posteriors of of $\{r_k\}$.
\end{itemize}

\subsection{Approximate and Exact Inference}

Although all proposed NB process models have closed-form conditional posteriors, they contain countably infinite atoms that are infeasible to explicitly represent  in practice. This infinite dimensional problem can be addressed by using a discrete base measure with $K$ atoms, i.e., truncating the total number of atoms to be $K$, and then doing Bayesian inference via block Gibbs sampling 
\cite{ishwaran2000markov}. This is a very general approach and is used in our experiments to make a fair comparison between a wide variety of  models. Block gibbs sampling for the Gamma-NB process is described  in Appendix B; 
block gibbs sampling for other NB processes and related algorithms in Table \ref{Tab:Relationships} can be similarly derived, as described in  \cite{MingyuanNIPS2012} and
omitted here for brevity. The infinite dimensional problem can also be addressed by discarding the atoms with weights smaller than a small constant $\epsilon$ \cite{Wolp:Clyd:Tu:2011} or by modifying the L\'{e}vy measure to make its integral over the whole space be finite \cite{BNBP_PFA}. A sufficiently large (small) $K$ ($\epsilon$) usually provides a good approximation, however, there is an increasing risk of wasting 
computation as the truncation level gets larger. 

To avoid truncation, the  slice sampling scheme of \cite{SliceSampling} has been utilized for the Dirichlet process and normalized random measure based mixture models \cite{papaspiliopoulos2008retrospective,walker2007sampling,griffin2011posterior}. 
With  auxiliary slice latent variables introduced to allow adaptive truncations in each MCMC interaction, the infinite dimensional problem is transformed into a finite one. This method has also been applied to the beta-Bernoulli process \cite{tehSB} and the beta-NB process \cite{NBPJordan}. It would be interesting to investigate slice sampling for the NB process based count and mixture models, which provide likelihoods that might be more amenable to posterior simulation since no normalization is imposed on the weights of the atoms. 
As slice sampling 
 is not the focus of this paper, we leave it for future study.

Both the block Gibbs sampler  and the slice sampler explicitly represent a finite set of atoms for posterior simulation, and algorithms based on these samplers are commonly referred as ``conditional'' methods \cite{papaspiliopoulos2008retrospective,kalli2011slice}. Another approach of solving the infinite dimensional problem is employing a collapsed inference scheme that marginalizes out the atoms and their weights \cite{Escobar1995,MullerDP,neal2000markov,Teh2010a,HDP,InfGaP}.  Algorithms based on the collapsed inference scheme are usually referred as ``marginal'' methods \cite{papaspiliopoulos2008retrospective,kalli2011slice}. A well-defined prediction rule is usually required to develop a  collapsed Gibbs sampler, and the conjugacy between the likelihood and the prior distribution of atoms is desired to avoid 
numerical integrations. In topic modeling, a word is linked to a Dirichlet distributed atom with a multinomial likelihood, thus the atoms can be analytically marginalized out; since their weights can also be marginalized out as in (\ref{eq:GammaNBPolyaUrn}), 
we may develop a collapsed Gibbs sampler for the gamma-NB process based topic models. As the collapsed inference scheme is not the focus of this paper and the prediction rules for other NB processes need further investigation, we leave them for future study.

\section{Example Results and Discussions}\label{sec:Results}

Motivated by  Table \ref{Tab:Relationships}, we consider topic modeling using a variety of  NB processes. 
We compare them with LDA \cite{LDA,FindSciTopic} and CRF-HDP \cite{HDP,HDP-HMM}, in which the latent count $n_{jk}$ is marginally distributed as 
\beq\label{eq:betabino}
n_{jk}\sim\mbox{Beta-Binomial}(N_j,\alpha \tilde{r}_k,\alpha(1-\tilde{r}_k))\notag
\eeq
with $\tilde{r}_k$ fixed as $1/K$ in LDA and learned from the data in CRF-HDP. For fair comparison, they are all implemented with block Gibbs sampling using a discrete base measure with $K$ atoms, and for the first fifty iterations, the Gamma-NB process with $r_k\equiv 50/K$ and $p_j\equiv 0.5$ is used for initialization. We set $K$ large enough that only a subset of the $K$ atoms would be used by the data. We consider 2500 Gibbs sampling iterations and collect the last 1500 samples.  

We consider the Psychological Review\footnote{\label{footnote}http://psiexp.ss.uci.edu/research/programs$\_$data/toolbox.htm} corpus, restricting the vocabulary to terms that occur in
five or more documents. The corpus includes 1281 abstracts from 1967 to 2003, with $V=2566$ and 71,279 total word counts.
We randomly select $20\%$, $40\%$, $60\%$ or $80\%$ of the words from each document  to learn a document dependent probability for each term $v$
and calculate the per-word perplexity on the held-out words as
 \beqs &\mbox{Perplexity} = \exp\left( - \frac{1}{y_{\cdotv\cdotv}}\sum_{j=1}^{J}\sum_{v=1}^V y_{jv} \log f_{jv}\right),
 \eeqs
 where $f_{jv} = \frac{\sum_{s=1}^S\sum_{k=1}^K\phi^{(s)}_{vk}\theta^{(s)}_{jk}}{\sum_{s=1}^S\sum_{v=1}^V\sum_{k=1}^K\phi^{(s)}_{vk}\theta^{(s)}_{jk}}$, $y_{jv}$ is the number of  words held out at term $v$ in document $j$, $y_{\cdotv\cdotv}=\sum_{j=1}^{J}\sum_{v=1}^V y_{jv}$, 
 and $s=1,\cdots,S$ are the indices of collected samples. Note that the per-word perplexity is equal to $V$ if $f_{jv}=\frac{1}{V}$, thus it should be no greater than $V$ for a topic model that works appropriately. 
The final results are averaged over five random training/testing partitions. The performance measure is  the same as the one used in \cite{BNBP_PFA} and similar to those in \cite{AsuWelSmy2009a,wallach09,VBHDP}.

Note that the perplexity per held-out word is a fair metric to compare topic models. 
 As analyzed in Section \ref{sec:PFA}, 
 NB process topic models can also be considered as factor analysis of the term-document count matrix under the Poisson likelihood, with $\phiv_k$ as the $k$th factor that sums to one and $\theta_{jk}$ as the factor score of the $j$th document on the $k$th factor, which can be further linked to other discrete latent variable models. 
 If except for proportions $\tilde{\thetav}_j$ and $\tilde{\rv}$, the absolute values, e.g., $\theta_{jk}$, $r_k$ and $p_k$, are also of interest, 
then the NB process based count and mixture models would be more appropriate than the Dirichlet process based mixture models.

\begin{figure*}[!tb]
\begin{center}
\includegraphics[width=2.0\columnwidth]
{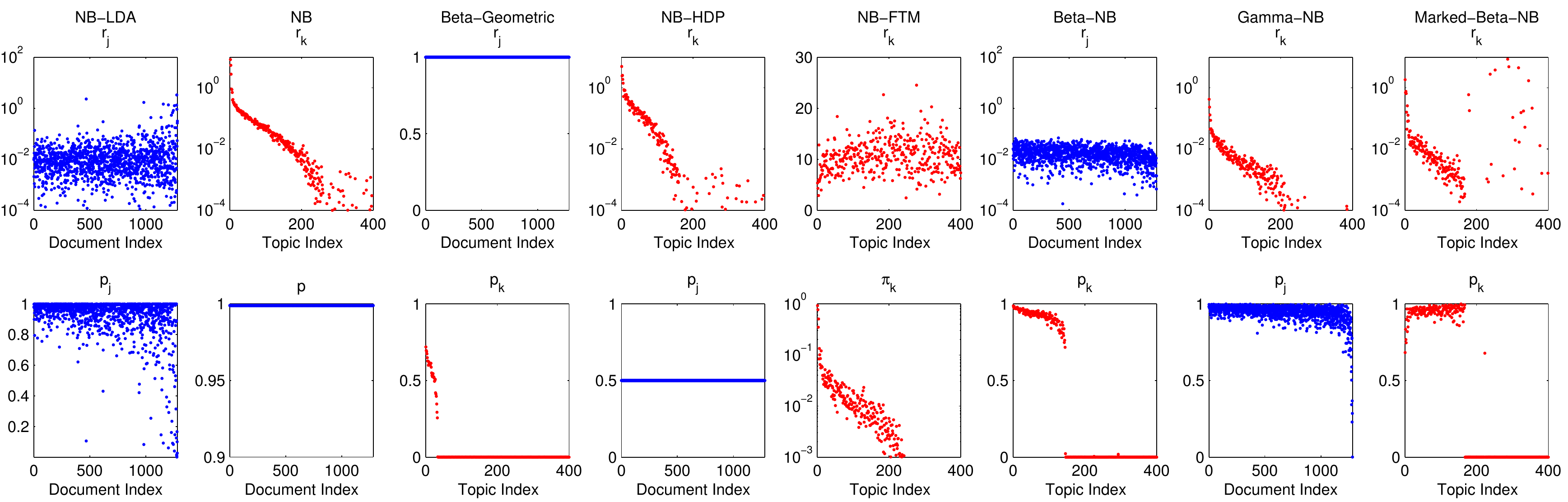}
\end{center}
\caption{ \label{fig:MakeSense}
Distinct sharing mechanisms and model properties are evident between various NB process topic models, by comparing their inferred NB dispersion parameters ($r_k$ or $r_j$) and probability parameters ($p_k$ or $p_j$). Note that the transition between active and non-active topics is very sharp when $p_k$ is used and much smoother when $r_k$ is used.  Both the documents and topics are ordered in a decreasing order based on the associated number of words. These results are based on the last MCMC iteration, on the Psychological Review corpus with 80\% of the words in each document used as training.
The values along the vertical axis are shown in either linear or log scales for convenient visualization. Document-specific and topic-specific parameters are shown in blue and red colors, respectively.
}
\end{figure*}

\begin{figure*}[!tb]
\begin{center}
\includegraphics[width=2.0\columnwidth]
{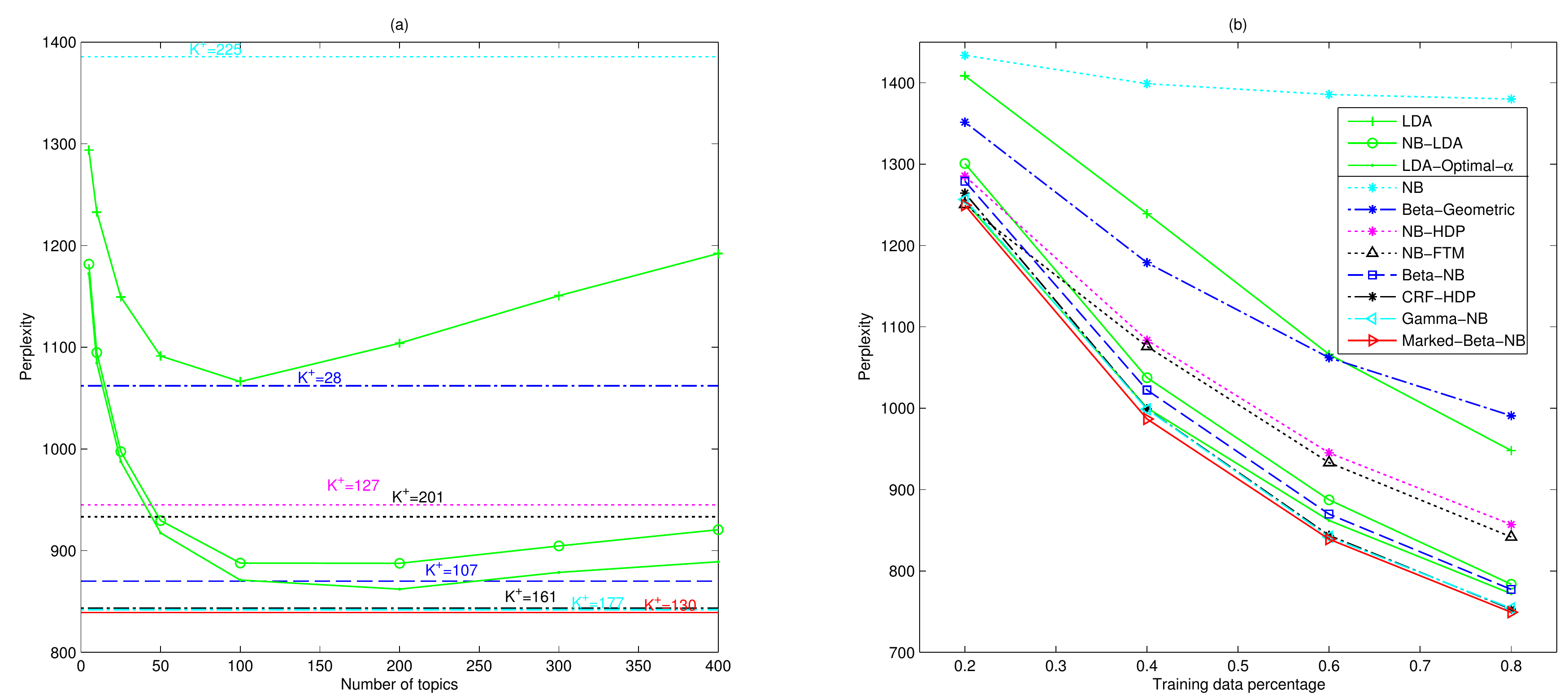}
\end{center}
\caption{ \label{fig:Perplexity}
Comparison of per-word perplexity on held out words between various algorithms listed in Table \ref{Tab:Relationships} on the Psychological Review corpus. LDA-Optimal-$\alpha$ refers to an LDA algorithm whose topic proportion Dirichlet concentration parameter $\alpha$ is optimized based on the results of the CRF-HDP on the same dataset.  (a) With $60\%$ of the words in each document used for training, the performance varies as a function of $K$ in both LDA and NB-LDA, which are parametric models, whereas the NB, Beta-Geometric, NB-HDP, NB-FTM, Beta-NB, CRF-HDP, Gamma-NB and Marked-Beta-NB all infer the number of active topics, which are 225, 28, 127, 201, 107, 161, 177 and 130, respectively, according to the last Gibbs sampling iteration. (b) Per-word perplexities of various algorithms as a function of the percentage of words in each document used for training. The results of LDA and NB-LDA are shown with the best settings of $K$ under each training/testing partition. Nonparametric Bayesian  algorithms listed in Table \ref{Tab:Relationships} are ranked in the legend from top to bottom according to their overall performance.
}
\end{figure*}

We show in Fig. \ref{fig:MakeSense} the NB dispersion and probability parameters learned  by various NB process topic models listed in Table \ref{Tab:Relationships}, revealing distinct sharing mechanisms and model properties. In Fig. \ref{fig:Perplexity} we compare the per-held-out-word prediction performance of various algorithms.
We set the parameters as $c=1$, $\eta=0.05$ and $a_0=b_0=e_0=f_0=0.01$.
For LDA and NB-LDA, we search $K$ for optimal performance. All the other NB process topic models are nonparametric Bayesian models that can automatically learn the number of active topics $K^+$ for a given corpus. For fair comparison, all the models considered are implemented with block Gibbs sampling, where $K=400$ is set as an upper-bound.

When $\theta_{jk} \equiv r_k$ is used, as in the NB process, different documents are imposed to have the same topic weights, leading to the worst held-out-prediction performance.

With a symmetric Dirichlet prior $\mbox{Dir}(\alpha/K,\cdots,\alpha/K)$ placed on the topic proportion for each document, the parametric LDA is found to be sensitive to both the  number of topics $K$ and the value of the concentration parameter $\alpha$. We consider $\alpha=50$, following the suggestion of the topic model toolbox 
provided for \cite{FindSciTopic}; we also consider an optimized value as $\alpha=2.5$, based on the results of the CRF-HDP on the same dataset.
As shown in Fig. \ref{fig:Perplexity}, when the number of training words is small, with optimized $K$ and $\alpha$, the parametric LDA can approach the performance of the nonparametric CRF-HDP; as the number of training words increases, the advantage of learning $\tilde{r}_k$ in the CRF-HDP than fixing $\tilde{r}_k=1/K$ in LDA becomes clearer. The concentration parameter $\alpha$ is important for both LDA and CRF-HDP since it controls the VMR of the count $n_{jk}$, which is equal to $(1-\tilde{r}_k)(\alpha+N_j)/(\alpha+1)$  based on (\ref{eq:betabino}). Thus fixing $\alpha$ may lead to significantly under- or over-estimated variations and then degraded performance, e.g., LDA with $\alpha=50$ performs much worse than LDA-Optima-$\alpha$, as shown in Fig. \ref{fig:Perplexity}. 

When $(r_j,p_j)$ is used, as in NB-LDA, different documents are weakly coupled with $r_j\sim\mbox{Gamma}(\gamma_0,1/c)$,
and the modeling results in Fig. \ref{fig:MakeSense} show that  a typical document in this corpus usually has a small $r_j$ and a large $p_j$, thus a large overdispersion level (ODL) and a large variance-to-mean ratio (VMR), indicating highly overdispersed counts on its topic usage. NB-LDA is a parametric topic model that requires tuning the number of topics $K$. It improves over LDA in that it only has to tune $K$, whereas LDA has to tune both $K$ and $\alpha$. With an appropriate $K$, the parametric NB-LDA may outperform the nonparametric NB-HDP and NB-FTM as the training data percentage increases, showing that even by learning both the NB 
parameters $r_j$ and $p_j$ in a document dependent manner, we may get better data fitting than using nonparametric models that 
fix the NB probability parameters.

When $(r_j,p_k)$ is used to model the latent counts $\{n_{jk}\}_{j,k}$, as in the Beta-NB process, the transition between active and non-active topics is very sharp that $p_k$ is either far from zero or almost zero, as shown in Fig.~\ref{fig:MakeSense}. That is because $p_k$ controls the mean  $\E[\sum_{j}n_{jk}]=p_k/(1-p_k)\sum_{j}r_j$ and the VMR  $(1-p_k)^{-1}$ on topic $k$, thus a popular topic must also have large $p_k$ and hence large overdispersion measured by the VMR; since the counts $\{n_{jk}\}_{j}$ are  usually overdispersed, particularly  true in this corpus, a small $p_k$ indicating a small mean and small VMR is not favored 
and thus is rarely observed.

The Beta-Geometric process is a special case of the Beta-NB process that $r_j\equiv 1$, which is more than ten times larger than the values inferred by the Beta-NB process on this corpus, as shown in Fig. \ref{fig:MakeSense}; therefore, to fit the mean $\E[\sum_{j}n_{jk}]=Jp_k/(1-p_k)$, it has to use a substantially underestimated $p_k$, leading to severely underestimated variations and thus degraded performance, as confirmed by comparing the curves of the Beta-Geometric and Beta-NB processes in Fig. \ref{fig:Perplexity}.

 When $(r_k,p_j)$ is used, as in the Gamma-NB process, the transition is much smoother that $r_k$ gradually decreases, as shown in Fig.~\ref{fig:MakeSense}. The reason is that  $r_k$ controls the mean $\E[\sum_{j}n_{jk}]=r_k\sum_{j}p_j/(1-p_j)$ and the ODL $r_k^{-1}$ on topic $k$, thus popular topics must also have large $r_k$ and hence small overdispersion measured by the ODL, and unpopular topics are modeled with small $r_k$ and hence large overdispersion, allowing rarely and lightly used topics. Therefore, we can expect that $(r_k,p_j)$ would allow more topics than $(r_j,p_k)$,  as confirmed in Fig. \ref{fig:Perplexity} (a) that the Gamma-NB process learns 177 active topics, obviously  more than the 107 ones of the Beta-NB process. With these analysis, we can conclude that the mean and the amount of overdispersion (measure by the VMR or ODL) for the usage of topic $k$ is positively correlated under $(r_j,p_k)$ and negatively correlated under $(r_k,p_j)$.

 The NB-HDP is a special case of the Gamma-NB process that $p_j\equiv 0.5$. From a mixture modeling viewpoint, fixing $p_j = 0.5$ is a natural choice as $p_j$ appears irrelevant after normalization. However, from a count modeling viewpoint, this would make a restrictive assumption that each count vector $\{n_{jk}\}_{k=1,K}$ has the same VMR of 2. It is also interesting to examine (\ref{eq:alphaPost}), which can be viewed as the  concentration parameter $\alpha$  in the HDP, allowing the adjustment of $p_j$ would allow a more flexible model assumption on the amount of variations between the topic proportions, and thus potentially better data fitting.

 The CRF-HDP and Gamma-NB process have very similar performance on predicting held-out words, although they have distinct assumption on count modeling: $n_{jk}$ is modeled as an NB distribution in the Gamma-NB process while it is modeled as a beta-binomial distribution in the CRF-HDP. The Gamma-NB process adjust both $r_k$ and $p_j$ to fit the NB distribution, whereas the CRF-HDP learns both $\alpha$ and $\tilde{r}_k$ to fit the beta-binomial distribution. The concentration parameter $\alpha$ controls the VMR of the count $n_{jk}$ as shown in (\ref{eq:betabino}), and we find through experiments that prefixing its value may substantially degrade the performance of the CRF-HDP, thus this option is not considered in the paper and we exploit the CRF metaphor to update $\alpha$ as in \cite{HDP,HDP-HMM}.


When $(r_k, \pi_k)$ is used, as in the NB-FTM model, our results in Fig.~\ref{fig:MakeSense} show that we usually have a small $\pi_k$ and a large $r_k$, indicating topic $k$ is sparsely used across the documents but once it is used, the amount of variation on usage is small. This 
 property might be helpful when there are excessive number of zeros that might not be well modeled by the NB process alone. In our experiments, the more direct approaches of using $p_k$ or $p_j$ generally yield better results, which might not be the case when excessive number of zeros could be better explained with the 
 beta-Bernoulli processes, e.g., when the training words are scarce, the NB-FTM can approach the performance of the Marked-Beta-NB process.

When $(r_k,p_k)$ is used, as in the Marked-Beta-NB process, more diverse combinations of mean and overdispersion would be allowed as both $r_k$ and $p_k$ are now responsible for the mean $\E[\sum_{j}n_{jk}]=Jr_kp_k/(1-p_k)$. As observed in Fig.~\ref{fig:MakeSense}, there could be not only large mean and small overdispersion (large $r_k$ and small $p_k$), indicating a popular topic frequently used by most of the documents, but also large mean and large overdispersion (small $r_k$ and large $p_k$), indicating a topic heavily used in a relatively small percentage of documents. Thus $(r_k,p_k)$ may combine the advantages of using only $r_k$ or $p_k$ to model topic $k$, as confirmed by the superior performance of the Marked-Beta-NB process. 

\section{Conclusions}
We propose a variety of negative binomial (NB) processes for count modeling,
which can be naturally applied for the seemingly disjoint problem of mixture modeling. The proposed NB processes are completely random measures, which assign independent random variables to disjoint Borel sets of the measure space, as opposed to Dirichlet processes, whose measures on disjoint Borel sets are negatively correlated. 
We reveal connections between various discrete distributions and discover unique data augmentation and marginalization methods for the NB process, with which we are able to unite count and mixture modeling, analyze fundamental model properties, and derive efficient Bayesian inference. 
We demonstrate that the NB process and the gamma-NB process can be recovered from the Dirichlet process and the HDP, respectively. We show in detail the theoretical, structural and computational advantages of the NB process. We examine the distinct sharing mechanisms and model properties of various NB processes, with connections made to existing discrete latent variable models under the Poisson factor analysis framework. Experimental results on topic modeling show the importance of modeling both the NB dispersion and probability parameters, which respectively govern the overdispersion level and variance-to-mean ratio for count modeling.

\ifCLASSOPTIONcompsoc
  \section*{Acknowledgments}
  \else
  \section*{Acknowledgment}
\fi

The authors would like to thank the two anonymous
reviewers and the editor for their constructive comments  that help improve the manuscript. 

%

\ifCLASSOPTIONcaptionsoff
  \newpage
\fi

\bibliography{References052013}

\begin{thebibliography}{10}

\bibitem{InvGaussianPoisson}
C.~Dean, J.~F. Lawless, and G.~E. Willmot.
\newblock A mixed {P}oisson-inverse-{G}aussian regression model.
\newblock {\em Canadian Journal of Stat.}, 1989.

\bibitem{LGNB}
M.~Zhou, L.~Li, D.~Dunson, and L.~Carin.
\newblock Lognormal and gamma mixed negative binomial regression.
\newblock In {\em ICML}, 2012.

\bibitem{JLS07}
J.~O. Lloyd-Smith.
\newblock Maximum likelihood estimation of the negative binomial dispersion
  parameter for highly overdispersed data, with applications to infectious
  diseases.
\newblock {\em PLoS ONE}, 2007.

\bibitem{Hofmann99probabilisticlatent}
T.~Hofmann.
\newblock Probabilistic latent semantic analysis.
\newblock In {\em UAI}, 1999.

\bibitem{LDA}
D.~Blei, A.~Ng, and M.~Jordan.
\newblock Latent {D}irichlet allocation.
\newblock {\em J. Mach. Learn. Res.}, 2003.

\bibitem{CannyGaP}
J.~Canny.
\newblock Gap: a factor model for discrete data.
\newblock In {\em SIGIR}, 2004.

\bibitem{DCA}
W.~Buntine and A.~Jakulin.
\newblock Discrete component analysis.
\newblock In {\em Subspace, Latent Structure and Feature Selection Techniques}.
  Springer-Verlag, 2006.

\bibitem{BNBP_PFA}
M.~Zhou, L.~Hannah, D.~Dunson, and L.~Carin.
\newblock Beta-negative binomial process and {P}oisson factor analysis.
\newblock In {\em AISTATS}, 2012.

\bibitem{PoissonP}
J.~F.~C. Kingman.
\newblock {\em Poisson Processes}.
\newblock Oxford University Press, 1993.

\bibitem{Wolpert98poisson/gammarandom}
R.~L. Wolpert and K.~Ickstadt.
\newblock Poisson/gamma random field models for spatial statistics.
\newblock {\em Biometrika}, 1998.

\bibitem{InfGaP}
M.~K. Titsias.
\newblock The infinite gamma-{P}oisson feature model.
\newblock In {\em NIPS}, 2008.

\bibitem{Thibaux}
R.~J. Thibaux.
\newblock {\em Nonparametric {B}ayesian Models for Machine Learning}.
\newblock PhD thesis, UC Berkeley, 2008.

\bibitem{Miller}
K.~T. Miller.
\newblock {\em Bayesian Nonparametric Latent Feature Models}.
\newblock PhD thesis, UC Berkeley, 2011.

\bibitem{NBPJordan}
T.~Broderick, L.~Mackey, J.~Paisley, and M.~I. Jordan.
\newblock Combinatorial clustering and the beta negative binomial process.
\newblock {\em arXiv:1111.1802v3}, 2012.

\bibitem{MingyuanNIPS2012}
M.~Zhou and L.~Carin.
\newblock Augment-and-conquer negative binomial processes.
\newblock In {\em NIPS}, 2012.

\bibitem{ferguson73}
T.~Ferguson.
\newblock A {B}ayesian analysis of some nonparametric problems.
\newblock {\em The Annals of Statistics}, 1973.

\bibitem{DP_Mixture_Antoniak}
C.~E. Antoniak.
\newblock Mixtures of {D}irichlet processes with applications to {B}ayesian
  nonparametric problems.
\newblock {\em Ann. Statist.}, 1974.

\bibitem{Escobar1995}
M.~D. Escobar and M.~West.
\newblock Bayesian density estimation and inference using mixtures.
\newblock {\em JASA}, 1995.

\bibitem{MullerDP}
S.~N. MacEachern and P.~M{\"u}ller.
\newblock Estimating mixture of {D}irichlet process models.
\newblock {\em Journal of Computational and Graphical Statistics}.

\bibitem{NealDPM}
R.~M. Neal.
\newblock Markov chain sampling methods for {D}irichlet process mixture models.
\newblock {\em JCGS}, 2000.

\bibitem{Teh2010a}
Y.~W. Teh.
\newblock {D}irichlet processes.
\newblock In {\em Encyclopedia of Machine Learning}. Springer, 2010.

\bibitem{Wolp:Clyd:Tu:2011}
R.~L. Wolpert, M.~A. Clyde, and C.~Tu.
\newblock Stochastic expansions using continuous dictionaries: {L\'e}vy
  {A}daptive {R}egression {K}ernels.
\newblock {\em Annals of Statistics}, 2011.

\bibitem{lijoi2007controlling}
A.~Lijoi, R.~H. Mena, and I.~Pr{\"u}nster.
\newblock Controlling the reinforcement in {B}ayesian non-parametric mixture
  models.
\newblock {\em Journal of the Royal Statistical Society: Series B}, 2007.

\bibitem{HDP}
Y.~W. Teh, M.~I. Jordan, M.~J. Beal, and D.~M. Blei.
\newblock Hierarchical {D}irichlet processes.
\newblock {\em JASA}, 2006.

\bibitem{HDP-HMM}
E.~Fox, E.~Sudderth, M.~Jordan, and A.~Willsky.
\newblock Developing a tempered {HDP-HMM} for systems with state persistence.
\newblock {\em MIT LIDS, TR \#2777}, 2007.

\bibitem{VBHDP}
C.~Wang, J.~Paisley, and D.~M. Blei.
\newblock Online variational inference for the hierarchical {D}irichlet
  process.
\newblock In {\em AISTATS}, 2011.

\bibitem{pitman1997two}
J.~Pitman and M.~Yor.
\newblock The two-parameter {P}oisson-{D}irichlet distribution derived from a
  stable subordinator.
\newblock {\em The Annals of Probability}, 1997.

\bibitem{ishwaran2001gibbs}
H.~Ishwaran and L.~F. James.
\newblock Gibbs sampling methods for stick-breaking priors.
\newblock {\em JASA}, 2001.

\bibitem{Kingman}
J.~F.~C. Kingman.
\newblock Completely random measures.
\newblock {\em Pacific Journal of Mathematics}, 1967.

\bibitem{JordanCRMbook}
M.~I. Jordan.
\newblock Hierarchical models, nested models and completely random measures.
\newblock In M.-H. Chen, D.~Dey, P.~Mueller, D.~Sun, and K.~Ye, editors, {\em
  Frontiers of Statistical Decision Making and Bayesian Analysis: in Honor of
  James O. Berger}. New York: Springer, 2010.

\bibitem{JordanBP}
R.~Thibaux and M.~I. Jordan.
\newblock {Hierarchical beta processes and the Indian buffet process}.
\newblock In {\em AISTATS}, 2007.

\bibitem{Hjort}
N.~L. Hjort.
\newblock Nonparametric {B}ayes estimators based on beta processes in models
  for life history data.
\newblock {\em Ann. Statist.}, 1990.

\bibitem{ishwaran02}
H.~Ishwaran and M.~Zarepour.
\newblock Exact and approximate sum-representations for the {D}irichlet
  process.
\newblock {\em Can. J. Statist.}, 2002.

\bibitem{PolyaUrn}
D.~Blackwell and J.~MacQueen.
\newblock Ferguson distributions via {P}\'olya urn schemes.
\newblock {\em The Annals of Statistics}, 1973.

\bibitem{aldous:crp}
D.~Aldous.
\newblock Exchangeability and related topics.
\newblock In {\em Ecole d'Ete de Probabilities de Saint-Flour XIII}, pages
  1--198. Springer, 1983.

\bibitem{csp}
J.~Pitman.
\newblock {\em Combinatorial stochastic processes}.
\newblock Lecture Notes in Mathematics. Springer-Verlag, 2006.

\bibitem{Yule}
M.~Greenwood and G.~U. Yule.
\newblock An inquiry into the nature of frequency distributions representative
  of multiple happenings with particular reference to the occurrence of
  multiple attacks of disease or of repeated accidents.
\newblock {\em Journal of Royal Stat. Soc.}, 1920.

\bibitem{LogPoisNB}
M.~H. Quenouille.
\newblock A relation between the logarithmic, {P}oisson, and negative binomial
  series.
\newblock {\em Biometrics}, 1949.

\bibitem{johnson2005univariate}
N.~L. Johnson, A.~W. Kemp, and S.~Kotz.
\newblock {\em Univariate Discrete Distributions}.
\newblock John Wiley \& Sons, 2005.

\bibitem{NB_Fitting_53}
C.~I. Bliss and R.~A. Fisher.
\newblock Fitting the negative binomial distribution to biological data.
\newblock {\em Biometrics}, 1953.

\bibitem{Cameron1998}
A.~C. Cameron and P.~K. Trivedi.
\newblock {\em Regression Analysis of Count Data}.
\newblock Cambridge, UK, 1998.

\bibitem{WinkelmannCount}
R.~Winkelmann.
\newblock {\em Econometric Analysis of Count Data}.
\newblock Springer, Berlin, 5th edition, 2008.

\bibitem{NB_Bio_2008}
M.~D. Robinson and G.~K. Smyth.
\newblock Small-sample estimation of negative binomial dispersion, with
  applications to {SAGE} data.
\newblock {\em Biostatistics}, 2008.

\bibitem{NB_Pieters_1977}
E.~P. Pieters, C.~E. Gates, J.~H. Matis, and W.~L. Sterling.
\newblock Small sample comparison of different estimators of negative binomial
  parameters.
\newblock {\em Biometrics}, 1977.

\bibitem{NB_1984}
L.~J. Willson, J.~L. Folks, and J.~H. Young.
\newblock Multistage estimation compared with fixed-sample-size estimation of
  the negative binomial parameter $k$.
\newblock {\em Biometrics}, 1984.

\bibitem{LawlessNB87}
J.~F. Lawless.
\newblock Negative binomial and mixed {P}oisson regression.
\newblock {\em Canadian Journal of Statistics}, 1987.

\bibitem{ML_NB90}
W.~W. Piegorsch.
\newblock Maximum likelihood estimation for the negative binomial dispersion
  parameter.
\newblock {\em Biometrics}, 1990.

\bibitem{BiasMLE_NB}
K.~Saha and S.~Paul.
\newblock Bias-corrected maximum likelihood estimator of the negative binomial
  dispersion parameter.
\newblock {\em Biometrics}, 2005.

\bibitem{NBbayesian}
E.~T. Bradlow, B.~G.~S. Hardie, and P.~S. Fader.
\newblock Bayesian inference for the negative binomial distribution via
  polynomial expansions.
\newblock {\em Journal of Computational and Graphical Statistics}, 2002.

\bibitem{DILN}
J.~Paisley, C.~Wang, and D.~Blei.
\newblock The discrete infinite logistic normal distribution for
  mixed-membership modeling.
\newblock In {\em AISTATS}, 2011.

\bibitem{MingyuanSAM2010}
M.~Zhou, C.~Wang, M.~Chen, J.~Paisley, D.~Dunson, and L.~Carin.
\newblock Nonparametric {B}ayesian matrix completion.
\newblock In {\em IEEE Sensor Array and Multichannel Signal Processing
  Workshop}, 2010.

\bibitem{dHBP}
M.~Zhou, H.~Yang, G.~Sapiro, D.~B. Dunson, and L.~Carin.
\newblock Dependent hierarchical beta process for image interpolation and
  denoising.
\newblock In {\em AISTATS}, 2011.

\bibitem{DictTopic}
L.~Li, M.~Zhou, G.~Sapiro, and L.~Carin.
\newblock On the integration of topic modeling and dictionary learning.
\newblock In {\em ICML}, 2011.

\bibitem{BPFA2012}
M.~Zhou, H.~Chen, J.~Paisley, L.~Ren, L.~Li, Z.~Xing, D.~Dunson, G.~Sapiro, and
  L.~Carin.
\newblock Nonparametric {B}ayesian dictionary learning for analysis of noisy
  and incomplete images.
\newblock {\em IEEE TIP}, 2012.

\bibitem{FocusTopic}
S.~Williamson, C.~Wang, K.~A. Heller, and D.~M. Blei.
\newblock The {IBP} compound {Dirichlet} process and its application to focused
  topic modeling.
\newblock In {\em ICML}, 2010.

\bibitem{IBP}
T.~L. Griffiths and Z.~Ghahramani.
\newblock Infinite latent feature models and the {I}ndian buffet process.
\newblock In {\em NIPS}, 2005.

\bibitem{knowles}
D.~Knowles and Z.~Ghahramani.
\newblock Infinite sparse factor analysis and infinite independent components
  analysis.
\newblock In {\em Independent Component Analysis and Signal Separation}, 2007.

\bibitem{Paisley_ICML}
J.~Paisley and L.~Carin.
\newblock Nonparametric factor analysis with beta process priors.
\newblock In {\em ICML}, 2009.

\bibitem{Mingyuan09}
M.~Zhou, H.~Chen, J.~Paisley, L.~Ren, G.~Sapiro, and L.~Carin.
\newblock Non-parametric {B}ayesian dictionary learning for sparse image
  representations.
\newblock In {\em NIPS}, 2009.

\bibitem{Dunson05bayesianlatent}
D.~B. Dunson and A.~H. Herring.
\newblock Bayesian latent variable models for mixed discrete outcomes.
\newblock {\em Biostatistics}, 2005.

\bibitem{NMF}
D.~D. Lee and H.~S. Seung.
\newblock Algorithms for non-negative matrix factorization.
\newblock In {\em NIPS}, 2001.

\bibitem{ishwaran2000markov}
H.~Ishwaran and M.~Zarepour.
\newblock Markov chain {M}onte {C}arlo in approximate {D}irichlet and beta
  two-parameter process hierarchical models.
\newblock {\em Biometrika}, 2000.

\bibitem{SliceSampling}
P.~Damlen, J.~Wakefield, and S.~Walker.
\newblock Gibbs sampling for {B}ayesian non-conjugate and hierarchical models
  by using auxiliary variables.
\newblock {\em Journal of the Royal Statistical Society B}, 1999.

\bibitem{papaspiliopoulos2008retrospective}
O.~Papaspiliopoulos and G.~O. Roberts.
\newblock Retrospective {M}arkov chain {M}onte {C}arlo methods for {D}irichlet
  process hierarchical models.
\newblock {\em Biometrika}, 2008.

\bibitem{walker2007sampling}
S.~G. Walker.
\newblock Sampling the {D}irichlet mixture model with slices.
\newblock {\em Communications in Statistics Simulation and Computation}, 2007.

\bibitem{griffin2011posterior}
J.~E. Griffin and S.~G. Walker.
\newblock Posterior simulation of normalized random measure mixtures.
\newblock {\em JCGS}, 2011.

\bibitem{tehSB}
Y.~W. Teh, D.~G\"{o}r\"{u}r, and Z.~Ghahramani.
\newblock {Stick-breaking construction for the {I}ndian buffet process}.
\newblock In {\em AISTATS}, 2007.

\bibitem{kalli2011slice}
M.~Kalli, J.~E. Griffin, and S.~G. Walker.
\newblock Slice sampling mixture models.
\newblock {\em Statistics and computing}, 2011.

\bibitem{neal2000markov}
R.~M. Neal.
\newblock Markov chain sampling methods for {D}irichlet process mixture models.
\newblock {\em JCGS}, 2000.

\bibitem{FindSciTopic}
T.~L. Griffiths and M.~Steyvers.
\newblock Finding scientific topics.
\newblock {\em PNAS}, 2004.

\bibitem{AsuWelSmy2009a}
A.~Asuncion, M.~Welling, P.~Smyth, and Y.~W. Teh.
\newblock On smoothing and inference for topic models.
\newblock In {\em UAI}, 2009.

\bibitem{wallach09}
H.~M. Wallach, I.~Murray, R.~Salakhutdinov, and D.~Mimno.
\newblock Evaluation methods for topic models.
\newblock In {\em ICML}, 2009.

\end{thebibliography}
\bibliographystyle{unsrt}

\appendices

\section{Proof of Theorem \ref{lem:PoisLog}} 
\label{sec:Proof}
With the PMFs of both the NB and CRT distributions, the PMF of the joint distribution of counts $m$ and $l$ is 
$f_{M,L}(m,l|r,p) =  f_L(l|m,r)f_M(m|r,p)={\frac{\Gamma(r)|s(m,l)| r^l}{\Gamma(m+r)}} \frac{\Gamma(r+m)(1-p)^r p^m}{m!\Gamma(r)} = \frac{|s(m,l)|r^l(1-p)^r p^m}{m!}$, which is the same as (\ref{eq:PMF}).

Since $m\sim{\mbox{SumLog}}(l,p)$ is the summation of $l$ iid  $\mbox{Log}(p)$ random variables, its PGF becomes
$
C_{M}(z)=C_{U}^l(z)=
\left[{\ln(1-pz)}/{\ln(1-p)}\right]^l,~ |z|<{p^{-1}}.
$ 
With 
$[\ln(1+x)]^l = l!\sum_{n=l}^\infty{s(n,l)x^n}/{n!}$ and $|s(m,l)|=(-1)^{m-l}s(m,l)$ \cite{johnson2005univariate}, its PMF can be expressed as 
\beqs\label{eq:m_l_p}
&f_{M}(m|l,p) = \frac{C_{M}^{(m)}(0)}{m!} = \frac{ p^m l! |s(m,l)|}{m! [-\ln(1-p)]^l}.
\eeqs
Letting $l \sim {\mbox{Pois}}(-r\ln(1-p))$, the PMF of the joint distribution of counts $m$ and $l$ is
$f_{M,L}(m,l|r,p)=f_{M}(m|l,p)f_{L}(l|r,p)= \frac{ p^m l! |s(m,l)|}{m! [-\ln(1-p)]^l} \frac{(-r\ln(1-p))^l e^{r\ln(1-p)}}{l!} 
= \frac{|s(m,l)|r^l(1-p)^r p^m}{m!}$, which is the same as
(\ref{eq:PMF}).
\qed

\section{Block Gibbs Sampling for the Gamma-Negative Binomial Process}\label{sec:GNBPInf}
With $p_j\sim\mbox{Beta}(a_0,b_0)$, $\gamma_0\sim \mbox{Gamma}(e_0,1/f_0)$ and a discrete base measure as $G_0=\sum_{k=1}^K \frac{\gamma_0}{K}\delta_{\omega_k}$,
following Section \ref{sec:post}, block Gibbs sampling for (\ref{eq:GNBP_mixture0}) proceeds as
\beqs \label{eq:inference}
\hspace{-3mm}&\mbox{Pr}(z_{ji}=k|-) \propto F(x_{ji};\omega_{k})\theta_{jk},\notag\\
\hspace{-3mm}&(l_{jk}|-) \sim \mbox{CRT}(n_{jk},r_k),~(l'_{k}|-) \sim \mbox{CRT}\left(\sum_{j} l_{jk},\gamma_0/K\right),\notag\\
\hspace{-3mm}&(p_j|-)\sim \mbox{Beta}\left(a_0+N_j,b_0+ \sum_{k}r_k\right),\notag\\
\hspace{-3mm}&(\gamma_0|-)  \sim  \mbox{Gamma}  \left( e_0  +   \sum_{k}l'_{k}, \frac{1}{f_0  -  \ln( 1 - p')} \right),\notag\\
\hspace{-3mm}&(r_k|-) \sim \mbox{Gamma} \left( \gamma_0/K +  \sum_{j}  l_{jk}, \frac{1}{c - \sum_{j} \ln( 1 - p_j)}\right), 
\notag\\
\hspace{-3mm}&(\theta_{jk}|-) \sim \mbox{Gamma}(r_k+n_{jk},p_j),\notag\\
\hspace{-3mm}&p(\omega_k|-) \propto g_0(\omega_k)\prod_{z_{ji}=k}F(x_{ji};\omega_k) ,
\eeqs
where $p': = \frac{-\sum_j\ln(1-p_j)}{c-\sum_j\ln(1-p_j)}$.
Note that when $K\rightarrow \infty$, we have $(l'_{k}|-)=
\delta(\sum_{j}n_{jk}>0)$ and thus $\sum_k l'_k = K^{+}$.

\end{document}